# Charging of dielectric surfaces in contact with aqueous electrolyte – the influence of $CO_2$

Peter Vogel[1*], Nadir Möller[1], Muhammad Nawaz Qaisrani[1], Pravash Bista[2], Stefan A. L. Weber[2], Hans-Jürgen Butt[2], Benno Liebchen[3], Marialore Sulpizi[4] and Thomas Palberg[1]

[1] Institute of Physics, Johannes Gutenberg University, 55128 Mainz, Germany
[2] Max Planck Institute for Polymer Research, 55128 Mainz, Germany
[3] Institute for Condensed Matter Physics, Technische Universität Darmstadt, 64289 Darmstadt, Germany
[4] Department of Physics, Ruhr Universität Bochum, 44780 Bochum, Germany

**Abstract:** The charge state of dielectric surfaces in aqueous environments is of fundamental and technological importance. Here, we study the influence of dissolved molecular $CO_2$ on the charging of three, chemically different surfaces ($SiO_2$, Polystyrene, Perfluorooctadecyltrichlorosilane). We determine their charge state from electrokinetic experiments. We compare an ideal, $CO_2$-free reference system to a system equilibrated against ambient $CO_2$ conditions. In the reference system, the salt-dependent decrease of the magnitudes of $\zeta$-potentials follows the expectations for a constant charge scenario. In the presence of $CO_2$, the starting potential is lower by some 50%. The following salt-dependent decrease is weakened for $SiO_2$ and inverted for the organic surfaces. We show that screening and pH driven charge regulation alone cannot explain the observed effects. As additional cause, we tentatively suggest dielectric regulation of surface charges due to a diffusively adsorbed thin layer of molecular $CO_2$. The formation of such a dynamic layer even at the hydrophilic and partially ionized silica surfaces is supported by a minimal theoretical model and results from molecular simulations.

## INTRODUCTION

The charge state of dielectric surfaces in contact with aqueous media is important for many contemporary technological challenges. Examples range from desalination, ice nucleation, and fog harvesting to nervous conduction, chemotaxis and fluid transport.[1–6] Charging plays a central role for stabilization of colloidal dispersions.[7–10] Charging, however, depends strongly on the environmental conditions. Interest in the charging process of surfaces in contact with fluids is therefore motivated at the same time by the complexity of this fundamental process and the prospect of useful application. Knowledge of the charge state under ideal lab conditions is vital to develop a principle understanding of mechanisms underlying the surface properties. Conversely, the charge state under ambient conditions is crucial for their performance.[2,9] Here we focus on the influence of dissolved $CO_2$ on the $\zeta$-potential of dielectric surfaces which has been neglected in most previous studies of this quantity.

In contact with a fluid, practically all dielectric surfaces are charged, either through dissociation of ionogenic surface groups[11] or by adsorbed ions.[12,13] In nominally neutral polymer spheres dispersed in organic solvents, even individual charging events were discriminated.[14] The equilibrium charge densities, respectively the corresponding potentials, are conveniently accessed in electrokinetic experiments. Such electrokinetic potentials are typically very close to the diffuse layer potentials but smaller than the intrinsic surface potentials. Conductivity yields the effective conductivity charge, i.e. the number, $Z_{eff}$ of effectively mobile micro-ions.[15] $\zeta$-potentials of planar or spherical surfaces are obtained from electrokinetic mobilities μ (where **v** = μ **E** is the corresponding electrokinetic slip velocity in an applied electric field **E**), which are evaluated by established IUPAC protocols based on the standard electrokinetic model.[16–18] For a constant charge density, one expects a decrease of the mobility magnitude with increasing electrolyte concentration due to screening effects. This has been confirmed in numerous experiments at elevated concentrations of added salt, $c_s \geq 10^{-3}$ mol L$^{-1}$.[7] At salt concentration below $c < 10^{-3}$ mol L$^{-1}$, however, data appear to be conflicting. Some authors report the expected decrease of mobility magnitudes[19–22], some an increase,[23] some find a maximum or even a minimum followed by a maximum.[24] The underlying reason is not yet resolved.

Closer inspection shows that the qualitatively different behaviour may relate to differences in the salt-free starting point. Salt-free or deionized water may contain several types of dissolved gases.[10] $CO_2$ can react to carbonic acid. The solubility of $CO_2$ in deionized water[25–27] at ambient conditions (standard air at 25 °C) is $1.18 \times 10^{-5}$ molL$^{-1}$. The carbonate related chemical reactions in $CO_2$ saturated salt-free aqueous solution are:[25–32]

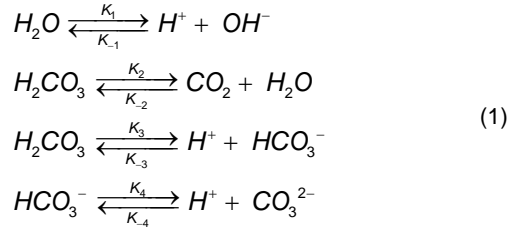

$$H_2O \underset{K_{-1}}{\overset{K_1}{\rightleftarrows}} H^+ + OH^-$$
$$H_2CO_3 \underset{K_{-2}}{\overset{K_2}{\rightleftarrows}} CO_2 + H_2O$$
$$H_2CO_3 \underset{K_{-3}}{\overset{K_3}{\rightleftarrows}} H^+ + HCO_3^-$$
$$HCO_3^- \underset{K_{-4}}{\overset{K_4}{\rightleftarrows}} H^+ + CO_3^{2-}$$

(1)

$K_i$ and $K_{-i}$ ($i = 1$–4) are the corresponding forward and backward kinetic constants. For the two-step acidic dissociation, p$K_{a1}$ = 6.35 and p$K_{a2}$ = 10.33.[33] At ambient conditions, they contribute 6.3 µmolL$^{-1}$ of electrolyte. Simultaneously, the pH is shifted to 5.5. A so-called realistic salt-free aqueous system results.[28–31] Under geologically relevant conditions of high pressures and high salt concentrations, the solubility of $CO_2$ is decreased due to salting-out.[27,34] Note that by contrast to other gases, like $N_2$ and $O_2$, the $CO_2$ content can be adjusted by ion exchange and transient exposition to ambient air. This renders the present experiments sensitive for the specific effects of $CO_2$.

The increased salinity alters the double layer structure, the screening parameter κ, and in turn also the electrokinetic behaviour.[28–30] The increased acidity alters the dissociation equilibrium of surface groups. Their degree of dissociation is determined by their p$K$ (e.g. $pK_a^{R-O-SO_3H} = 1.5$, and $pK_a^{R-COOH} = 4.5$, $pK_a^{SiOH} = 7.5$)[35] and the amount of protonic counter-ions accumulated close to the surface upon double layer formation.[31] A decrease in bulk pH enhances the proton accumulation and thus decreases the degree of dissociation (pH driven charge regulation).

Dissolved gases may further adsorb at surfaces.[10] Neutron reflectivity as well as x-ray measurements suggest the presence of a reduced (deuterated) water density region that depends on the type and concentration of dissolved gases.[36,37] Molecular dynamics simulations on a Lennard-Jones system report a significant enrichment of gas density close to hydrophobic surfaces.[38] Molecular hydrophobicity was discovered in spectroscopic experiments on overall hydrophilic silica.[39] Also experiments on colloidal suspensions, conducted at ambient pressure and µmolar salt concentrations suggest, that $CO_2$ can become stored at or on particle surfaces.[40,41] Gas has even been observed to form films or nano-bubbles at dielectric surfaces.[42,43] Being of immense practical importance, gas adsorption from solution and the equilibrium structure of gas-containing water near solid surfaces thus remain under intense investigation.[10,44] However, somewhat surprisingly, a systematic study on the influence of dissolved (and possibly accumulated) $CO_2$ on the charge state of dielectric surfaces and on its salt concentration dependence is still missing.

The present paper addresses this issue for three different dielectric surfaces. We study standard glass slides ($SiO_2$) as hydrophilic, inorganic material, slides coated by Perfluorooctadecyltrichlorosilane (PFOTS) as strongly hydrophobic sample, and 40:60 Poly-n-butylacrylamide-Polystyrene-copolymer (PnBAPS) spheres as mildly hydrophobic organic surface. All three surfaces are negatively charged: the glass bears terminal $OH^-$ groups, the perfluorinated surface is presumably charged by adsorbed $OH^{-\ 45}$, and the polymer spheres carry ionizable carboxyl and sulphate groups. Under ambient conditions, their contact angles decrease with increasing magnitude of the ζ-potential (Figure S4 of the supplementary materials). Experiments at $CO_2$-free conditions serve as reference. We compare the results for the reference case to those obtained at different levels of $CO_2$ and of added NaCl or HCl. By adding NaCl at equilibrium with airborne $CO_2$, we approach conditions typical for practical applications.

We mount the specimen slides as side walls of a specially designed electrophoretic cell [46] (Figure S5a of the supporting information). We fill the cell with a preconditioned suspension of PnBAPS tracer spheres, diluted to the desired low concentration. Cell dimensions and tracer concentrations are chosen such as to prevent double layer overlap. The cell is connected to a gas tight conditioning circuit comprising an ion exchange column to replace all ions in solution by $H^+$ and $OH^-$ (Figure S1 of the supporting information).[47] Electrolyte concentration and pH are monitored in-situ.[15,48] Our setup allows complete removal of dissolved and adsorbed $CO_2$ *via* its ionic reaction products. We stress that other dissolved gases, like $N_2$ or $O_2$ will not be affected by ion exchange. Starting from $CO_2$ equilibrated DI water, the pH increases from 5.5 to 7.0 and we obtain a stable conductivity of 55–60nScm$^{-1}$ within a few minutes. Starting from pre-conditioned suspensions, the removal of dissolved $CO_2$ takes about one hour for typical tracer volume fractions of $\Phi \approx 0.0005 - 0.001$. The final conductivity values are larger and with a small offset proportional to the tracer concentration.

Equilibration with airborne $CO_2$ starts from the thoroughly deionized state. The suspensions are cycled in contact with air until after about an hour, conductivity, pH, and $\zeta$-potential have attained constant values. Intermediate $CO_2$ concentrations are obtained for shorter contact times with ambient air.

The mobilities of spherical and flat surfaces are reliably determined by a recently introduced super-heterodyne version of laser Doppler velocimetry (SH-LDV),[49] which is particularly suited to determine electrokinetic velocities at transparent, planar specimens. It features a low angle integral measurement across the complete sample cross section. It records the distribution of tracer particle velocities $\mathbf{v}_p$ in an applied electric field, $\mathbf{E}$, in terms of their Doppler spectra, $C_{SH}(q,\omega)$. The spectral shape is given by a Lorentzian of width $\Gamma = \mathbf{q}^2 D_{eff}$ convoluted with the distribution $p(\mathbf{q} \cdot \mathbf{v}_p)$, where $\mathbf{q}$ is the scattering vector and $D_{eff}$ the effective tracer diffusion coefficient.[50] Super-heterodyning shifts the signal away from low-frequency noise and homodyne scattering. Figure 1a shows a field strength dependent series of spectra recorded for a salt-free system under $CO_2$-free reference conditions on PnBAPS tracers in a cell with PFOTS-coated walls.

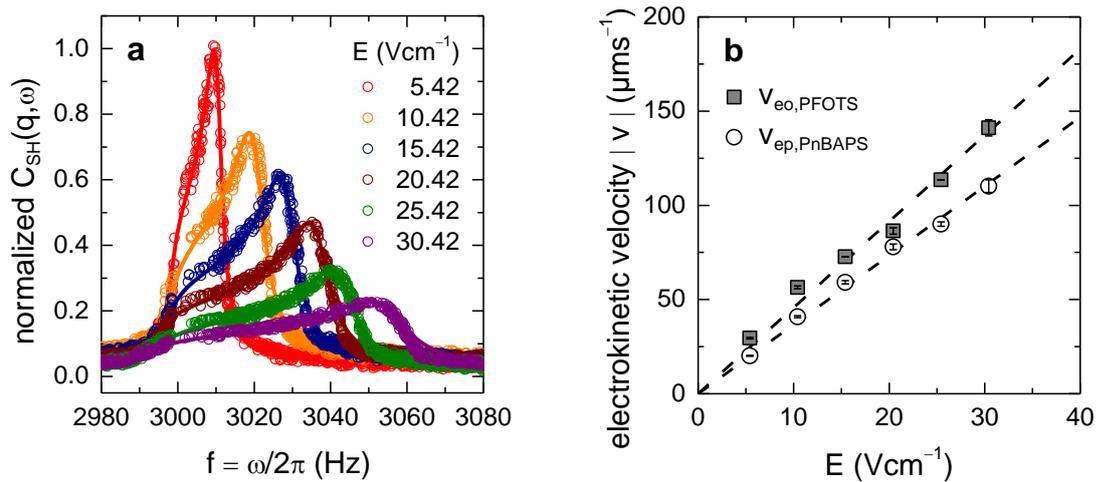

**Figure 1.** a) Doppler spectra (symbols) recorded on PnBAPS tracers in a cell with PFOTS-coated walls at salt and $CO_2$-free conditions. Data are for different field strengths as indicated in the key. Solid lines are least squares fits of Equation S1–S3. b) Moduli of electro-kinetic velocities plotted versus *E*. Error bars denote the standard error of the fits in a) at a confidence level of 0.95. Dashed lines are least squares fits of a linear function to the data.

The SH-frequency of $f_{SH} = \omega_{SH}/2\pi = 3$kHz corresponds to zero particle velocity. The characteristic spectral shape corresponds to a parabola like velocity profile.[51] With increasing field strength, the spectra stretch and their centre of mass shifts towards larger Doppler frequencies. The solid lines are least squares fits of our light scattering model (Equation S1–S4 of the Supplementary Materials). They return three independent parameters: The tracer velocity against the solvent, $\mathbf{v}_{ep} = \mu_{PnBAPS} \mathbf{E}$ from the spectral center of mass, the solvent velocity at the charged walls, $\mathbf{v}_{eo} = \mu_{PFOTS} \mathbf{E}$ from the field dependent width, and the tracer diffusivity, $D_{eff}$, from the homogeneous broadening.

Further details on materials, sample conditioning, experimental set-up and raw-data analysis are given in the supplementary material.

## RESULTS AND DISCUSSION

Figure 1b shows the moduli of velocities extracted from the spectra of Figure 1a. They increase linearly with field strength, as expected in the linear response regime. Least square fits of a linear function (dashed lines) return mobilities of $\mu_{PnBAPS} = -(3.1\pm0.2)\times10^{-8}$ $m^2V^{-1}s^{-1}$ and $\mu_{PFOTS} = -(4.0\pm0.1)\times10^{-8}\ m^2V^{-1}s^{-1}$, respectively. The quoted uncertainties correspond to the standard errors of the fits at a confidence level of 0.95. For $SiO_2$ at deionized and $CO_2$-free conditions, we obtained $\mu_{eo,SiO2} = -(9.0\pm0.1)\times10^{-8}\ m^2V^{-1}s^{-1}$ (Figure S7 of the supplementary materials). Note, that as expected, all three surfaces are negatively charged. As the electrokinetic model is symmetric upon charge reversal, we below mainly report the absolute values (magnitudes).

For all measurements, the mobilities are converted to reduced mobilities $\mu_{red} = 3\mu\eta_s e/2\varepsilon_0\varepsilon_r k_B T$.[16] These are further evaluated for $\zeta$-potentials using the standard electrokinetic model[17] adapted to treat realistic salt free conditions.[18] The model accounts for the systematic reduction in mobility magnitudes at spherical surfaces due to double layer relaxation effects not present for flat surfaces (cf. Figure S8 of the supporting materials). It thereby introduces a dependence of mobilities on the tracer size and the screening parameter $\kappa$. The latter is calculated from the concentrations of all micro-ionic species present in the system (particle counter ions, ions from water hydrolysis, ions from $CO_2$ reaction, and added salt ions). From the experiments at salt and $CO_2$-free conditions (Figure 1 and S7), we obtain: $\zeta_{PnBAPS} = -(61\pm4)$ mV, $\zeta_{PFOTS} = -(49\pm1)$ mV, and $\zeta_{SiO2} = -(112\pm1)$ mV, respectively.

In Figure 2a we compare the salt concentration-dependent mobility magnitudes for the three dielectric surfaces. Data are plotted against the total micro-ion concentration $c_s = \sum_i c_i = \sum_i n_i/(1000 N_A)$ summing over all micro-ionic species (for details of calculation and cross-check see supplementary information).[15,32] For all three data sets, the mobility magnitudes decreases monotonically with increasing electrolyte (NaCl) concentration. The corresponding magnitudes of the $\zeta$-potentials are shown in Figure 2c. Note that going from mobilities to potentials, the curves for PFOTS slides (grey squares) and PnBAPS spheres (open circles) switch place due to relaxation effects.

All three $\zeta$-potentials decrease in magnitude with added salt. At $c_s = 1000$ µmolL$^{-1}$, we obtain $\zeta_{SiO2} = -(83\pm1)$ mV, $\zeta_{PnBAPS} = -(35\pm2)$ mV, and $\zeta_{PFOTS} = -(36\pm1)$ mV. The corresponding slip-plane charge densities are $\sigma_{SiO2} = -(0.49\pm0.01)$ µCcm$^{-2}$, $\sigma_{PnBAPS} = -(0.14\pm0.02)$ µCcm$^{-2}$, and $\sigma_{PFOTS} = -(0.15\pm0.01)$ µCcm$^{-2}$ using Grahame´s equation.[35] Further, for all three surfaces in the semi-log plot of Figure 2c, the decrease in magnitude is approximately linear. This is theoretically expected for surfaces of constant surface charge subject to screening by increasing concentrations of added monovalent salt.

Next, we repeated the experiments under ambient conditions. Resulting mobility magnitudes are shown in Figure 2b, resulting magnitudes of $\zeta$-potentials in Figure 2d. In the presence of $CO_2$ but still without salt, all mobilities and $\zeta$-potentials are significantly lower than in the $CO_2$-free state. In fact, irrespective of material, they drop by about a factor of two. Further, upon the addition of salt, all three $\zeta$-potentials change in non-linear ways in this semi-log plot indicating a deviation from the constant charge scenario seen without $CO_2$. The salt concentration dependent decrease is significantly weakened for $SiO_2$, and for both organic surfaces, we even observe an increase in the magnitudes of the $\zeta$-potential with increasing salt concentration!

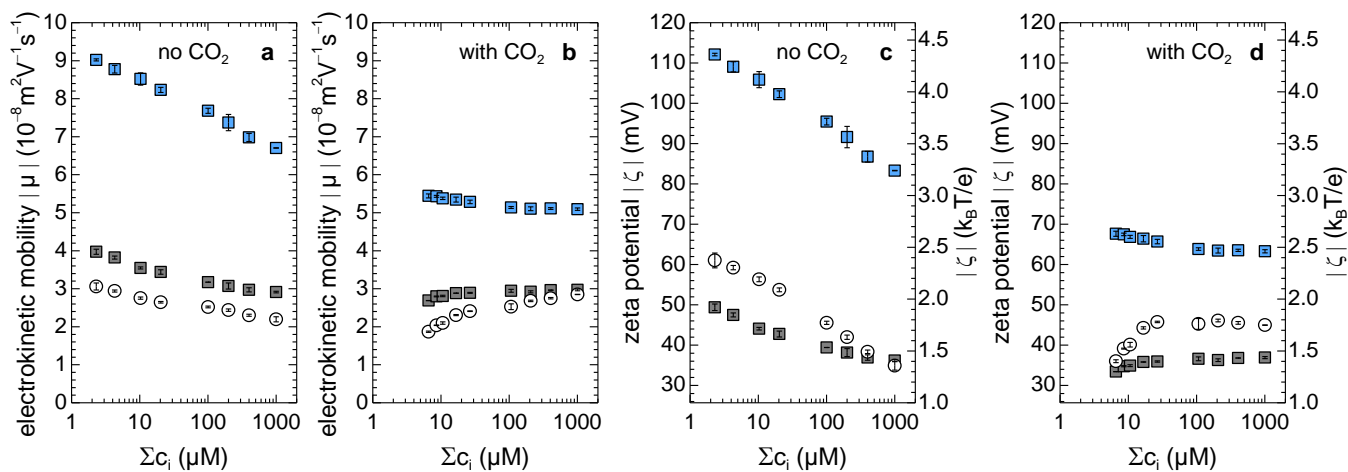

**Figure 2.** Comparison of the magnitudes of mobilities (left) and $\zeta$-potentials (right) observed without and with $CO_2$ upon addition of a 1:1 electrolyte (NaCl). Results for $SiO_2$ (blue squares), PnBAPS (open circles) and PFOTS-coated glass (grey squares) are plotted versus the logarithm of the total micro-ion concentration, including particle counterions, ions from water hydrolysis, added salt and $CO_2$ reaction products (Equation 1). Error bars denote standard errors of the linear least squares fits (see Figure 1b) at a confidence level of 0.95. Note that all three surfaces are negatively charged. Note further, that a) and c) correspond to pH ≈ 7 while b) and d) correspond to pH ≈ 5.5. a) Magnitudes of mobilities obtained for the $CO_2$ free reference system. b) Magnitudes of mobilities as obtained after $CO_2$ equilibration in contact with ambient air. c) Magnitudes of $\zeta$-potentials obtained for the reference system in mV (left scale) and reduced units (right scale). d) same as c) but for systems equilibrated in contact with ambient air.

Figure 2a confirms the theoretical expectation for $CO_2$ free systems, i.e., we find a decrease of mobility magnitudes with increasing salt concentration. Figure 2c shows that it relates to a reduction of the $\zeta$-potential magnitudes upon enhanced Double layer compression due to screening. Comparison to Figures 2b and d, however, unequivocally demonstrates a significant $CO_2$-induced effect for both spheres and flat surfaces, and, even more importantly for hydrophobic as well as hydrophilic surfaces. The strong drop in the absence of salt is present for three rather different dielectrics. The changes in the salt dependence differ in detail, but all three surfaces are strongly affected. We therefore exclude specific chemical reaction as underlying reason. Further, due to the employed $CO_2$-specific degassing process, we can exclude $N_2$ or $O_2$ to be responsible for the observed behaviour.

A decrease of magnitudes of the $\zeta$-potentials upon the addition of $CO_2$ can be expected from the corresponding increase in screening and from the decrease in pH. To discriminate between these two causes, we studied the $\zeta$-potentials of glass and PnBAPS as a function of overall electrolyte concentration for three different added chemicals. We started from a salt and $CO_2$ free state. In a first set of experiments, we added NaCl, which only contributes to screening. In a second set of experiments, we alternatively added HCl, which increases the electrolyte concentration, but also shifts the pH. In a third set of experiments, we again started from a salt free reference state, but now we altered the $CO_2$ content under conductivity and pH monitoring. Both latter quantities gave consistent values for the electrolyte concentration (see also supplementary information). The results for the different additives are displayed as a function of total micro-ion concentration in Figure 3a for $SiO_2$ and in Figure 3b for PnBAPS.

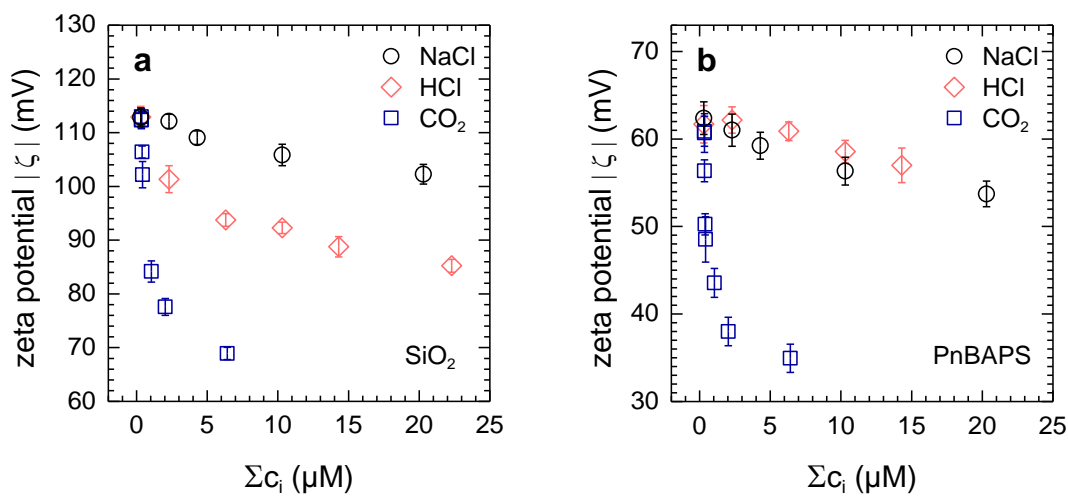

**Figure 3.** Magnitudes of $\zeta$-potentials of a) $SiO_2$ and b) PnBAPS in dependence on total micro-ion concentration including particle counterions, ions from water hydrolysis, salt (NaCl or HCl) and $CO_2$ reaction for three different added chemicals as indicated in the key. Note that for NaCl, the pH does not change. For HCl and $CO_2$ it varies in unison between very close to 7 to 5.5 for total micro-ion concentrations between 0.03μmol L$^{-1}$ at the start of the experiments (where only H$^+$ counterions and reaction products of water are present) and 6.3μmol L$^{-1}$ at the endpoint of the $CO_2$ (where either added H$^+$ and Cl$^-$ or formed $CO_2$-reaction products are present).

It is instructive, to compare the results at an electrolyte concentration of 6.3 µmolL$^{-1}$, which is the amount provided through carbonic acid in the absence of other electrolytes. Due to screening (NaCl), the magnitude of the $\zeta$-potential drops from (112±3) mV to (107±3) mV for glass and from (61±4) mV to (57±2) mV for PnBAPS. I.e., both surfaces become less negatively charged. The combined effects of screening and pH driven charge regulation (HCl) induce a drop to (94±2) mV for glass and (60±2) mV for PnBAPS. The glass surface with weakly acidic groups becomes even less charged than for NaCl. The PnBAPS surface bearing dominantly dissociated strongly acidic groups remains practically unchanged due to the small pH shift and shows only the effects of screening. A similar behaviour would be expected for the carbonic acid provided by dissolved $CO_2$. However, in the presence of $CO_2$ as only added electrolyte we observe a drop to (68±1) mV for $SiO_2$ and to (35±2) mV for PnBAPS. At the same time, the effective conductivity charge of the tracers drops from $Z_\sigma = 2350$ to $Z_\sigma = 1300$ (Figure S3b in the supplementary materials). For both our surfaces, the combined effects of increased electrolyte concentration and of pH shift are way too small to explain the observed factor two reduction upon addition of $CO_2$. We conclude that this additional drop must be due to the presence of molecular $CO_2$.

Neither decreased magnitudes of the $\zeta$-potentials upon the controlled addition of $CO_2$ nor their increase upon successive addition of NaCl have been reported before. Further, neither observation is backed by present theory assuming a simple double layer structure.[18] We therefore feel free to speculate. We tentatively suggest that the underlying reason might be some kind of restructuring of the innermost part of the double layer. We propose a diffusively adsorbed, thin but highly populated layer of molecular $CO_2$, which decreases the surface charge by *dielectric* charge regulation.

Weak acids dissociate in solution completely at neutral pH and infinite dilution. Similarly, the dissociation of surface groups is controlled by the local pH at the surface and the mutual electrostatic interaction between dissociated groups. The former was discussed in connection with bulk pH changes as well as with results from colloidal probe force microscopy[52] and the observation of charge reduction in strongly interacting colloidal suspensions.[31] It there was shown to be a double layer overlap-driven charge regulation. Regarding the latter, we have an average lateral separation of acidic groups of $d = 1.7$nm from titration. For groups to dissociate, the electrostatic interaction energy should not exceed the thermal energy. The distance at which Coulomb and thermal energy balance is known as Bjerrum length, $\lambda_B = e^2/4\pi\varepsilon_0\varepsilon_r k_B T$. In pure water, $\lambda_B \approx 7\text{Å}$. $\lambda_B$ increases with decreasing dielectric constant. Too closely spaced groups will dissociate less. In the absence of $CO_2$, we can assign a minimum area required for an electric charge as $A^+_{min} = \lambda_B^2 \approx 0.5\text{nm}^2$. This naturally limits the bare charge number for large group numbers $N$. This is corroborated by conductivity measurements under addition of NaCl in the absence of $CO_2$. Evaluated within Hessinger's model of independent ion migration, we obtain a bare charge of $Z \approx 1\times10^4$.

We now assume the surface groups to be located inside a thin diffuse layer of $CO_2$. Embedding the ionogenic groups in a diffuse layer of $CO_2$ decreases the dielectric constant and increase the electrostatic interaction between surface charges. We are aware that the exact value depends on the structure, shape, and extension of the layer as well as on the amount of $CO_2$ adsorbed. For illustration we assume $\varepsilon_r = 20$ for the average value in the immediate group environment. This gives $A^\pm_{min} \approx 7.8\text{nm}^2$ lowering the maximum charge density by a factor of 16. From titration of the PnBAPS spheres, the physical parking area of ionogenic groups is $A_0 \approx 3\text{ nm}^2 < A^+_{min,CO2}$. This is consistent with the observed reduction in $\zeta$-potential magnitudes of the three surfaces and of the PnBAPS diffuse layer charge. This conjecture predicts a larger charge reduction for surfaces with more highly populated diffuse $CO_2$-layer. Preliminary results from experiments using water equilibrated to atmospheres of differing $CO_2$ partial pressures including pure $CO_2$ point this way and will be published elsewhere. [PV, TP unpublished]

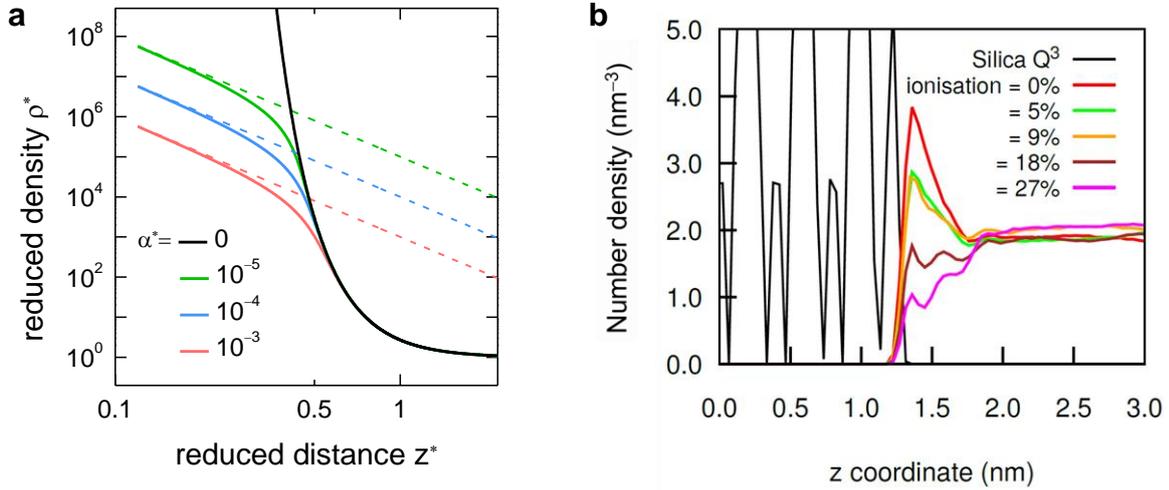

**Figure 4.** a) Reduced $CO_2$ density as a function of reduced distance. $z^* = 1$ corresponds to roughly one $CO_2$ LJ diameter. Solid curves are results for different values of $\alpha^*$, as indicated in the key. Dotted lines illustrate the power law limit for small distances. b) Density profile of $CO_2$ at Q3 surface with a silanol density of 4.7 $\text{nm}^{-2}$ and different degrees of ionisation.

The existence of a layer of neutral gas molecules at dielectric surfaces has been discussed in several experimental and theoretical studies.[10,36,38,40–44] We can support this idea for the specific case of $CO_2$ in a generic mean field model and by first results of molecular simulations based on established force fields. We first check, whether – in analogy to the formation of a Guoy-Chapman layer of counter-ions at charged surfaces – a $CO_2$- layer could form at all i.e., whether it is energetically feasible at all that effective $CO_2$-wall interactions of a conceivable strength (on the order of $k_B T$ at close contact) lead to a $CO_2$ enrichment layer with a thickness comparable to the size of the $CO_2$ molecules. In a minimal model, we consider a plane surface of uncharged bulk silica in contact with a structureless fluid of relative dielectric constant $\varepsilon_r = 80$. For the $CO_2$-molecules, we solved the $N$-particle Smoluchowski-equation for the joint probability density $P(\mathbf{r}_1, \mathbf{r}_2, \ldots, \mathbf{r}_N)$ in mean field with an isotropic effective interaction potential $U(r)$ (e.g. of Lennard Jones type (LJ)) in an external wall potential $U_{ext}(z) = -(f_0/3)/z^3$, representing effective Van der Waals attractions between a (thick) wall and the $CO_2$ molecules. We obtain the density distribution $\rho(z)$ (see supplementary information). In reduced units our result reads:

$$\rho^*(z^*) = \frac{1}{\alpha^*} W\left( \alpha^* \exp\left( \alpha^* + \frac{1}{z^{*3}} \right) \right) \qquad (2)$$

Here, $W$ is the product logarithm. We further use the reduced density, $\rho^* = \rho/\rho(z \to \infty)$ and the reduced distance $z^* = z/\xi$, where the characteristic length scale, $\xi = (f_0/(3k_B T))^{1/3}$, depends on the choice of the wall-$CO_2$ attraction strength $f_0$. Notably, we have only a single dimensionless control parameter, $\alpha^*$, for the shape of the curves, depending on the $CO_2$-$CO_2$ interaction strength, $\varepsilon$, the $CO_2$ saturation limit, $c_{max}$, and the thermal energy, $k_B T$, but not on the wall attraction, which enters only via the length scale $\xi$. Figure 4a shows exemplary results in terms of these reduced parameters for the case of a glass surface in contact with $CO_2$ saturated water (effective LJ interaction coefficient $\varepsilon = k_B T/2$, wall-$CO_2$ attraction strength $f_0 = 2.5 \times 10^{-49}$Jm$^3$). Here $z^* = 1$ corresponds to approximately the size of the $CO_2$ molecules. All three curves converge around $z^* = 0.5$, from whereon the curves decay in unison. Overall, from Figure 4a, we find that effective $CO_2$-wall attractions of a strength of less than $1 k_B T$ lead to a $CO_2$- enrichment zone with a thickness comparable to the size of the $CO_2$ molecules and which is rather insensitive to the strength (and type) of the effective $CO_2$-$CO_2$ cross interactions. The dashed lines are fits to the curves at small distances (see SI for details).

We conclude that already this minimal model supports the notion that there should be a thin, diffuse layer of significantly enhanced $CO_2$ density present at an uncharged dielectric wall. We explicitly note that this solely hinges on the assumption of a weak effective $CO_2$-wall attraction and does not consider any effects of hydrophobicity/philicity, wall charges and the corresponding competition with water adsorption.

To resolve these additional effects, we started molecular dynamics simulations employing realistic force fields and considered the different effects step by step. The silica Q4 surface is uncharged silanol-free silica, the silica Q3 surface may in addition carry silanol groups up to a density of 4.7 nm$^{-2}$.[53] For comparison we also considered l-Isoleucine as strictly hydrophobic surface using the standard Charmm27 force field.[54,55] Water was described by TIP3P model.[56] The $CO_2$ interaction were modelled following [57]. First, we compared water and $CO_2$ adsorption at the l-Isoleucine and the Q4 silica. While there is less water structure (layering) at the hydro-phobic surface, we see the same drastic accumulation of $CO_2$ in a single layer for both surfaces (Figure S9 and S10 of the supporting information). Next, we made the silica surfaces more hydrophilic, going from pure Q4 silica surfaces (oxygen bridging) to a mixture of Q4 and Q3 sites (silanol-bearing). With increasing silanol density, the amount of $CO_2$ adsorption is diminished, but still significant (Figure S11 of the supplementary materials). Even, if we allow for the tenfold larger silanol density than titrated, we still see a sevenfold enrichment of $CO_2$ density as compared to the bulk solution. This shows that even at strongly hydrophilic surfaces we find a pronounced $CO_2$ adsorption despite the competition with water layer formation. Finally, we allow for ionization using Sodium counter-ions. For an increasing degree of ionization, the adsorption of $CO_2$ is diminished. For a degree of deionization up to $\approx 9\%$ of a Q3 surface with a silanol density of 4.7 nm$^{-2}$ shown in Figure 4b (corresponding to a parking area for the charges of $A^\pm \approx 2.3$nm$^2$), there still is a clearly visible $CO_2$ enrichment. We note, that at such a charge density, we would already expect a substantial degree of charge regulation under conditions of an effectively decreased relative permittivity. Only for still larger degrees of dissociation, the effect is inverted, and $CO_2$ is expelled from the immediate surface region. This region is not reached in our experiments. The observation could, however, be relevant for surfaces with strongly acidic surface groups.

Our results strongly support the idea of the formation of a thin dynamic layer of $CO_2$ at hydrophobic and hydrophilic surfaces despite the presence of very large amounts of silanol groups and a moderate degree of ionization. These results are still preliminary, in the sense, that we have not yet implemented charge regulation. We anticipate that this has positive feedback on $CO_2$ layer formation, since the surface group interactions will be increased in the presence of $CO_2$, which in turn reduces the degree of ionization and allows for enhanced adsorption. Corresponding studies are under way and will be published elsewhere. These may further allow discussing the interesting results of Cyran et al.[39]

Not much can be said at this stage about our second main experimental finding, the increase in magnitude of the $\zeta$-potential with increasing salt concentration in the presence of $CO_2$, i.e., the re-charging. Several mechanisms could be anticipated, First, an increase of $CO_2$ solubility with increasing salinity. We deem this not very probable given the opposite effect of salting out at high salt concentration.[27,34] Second, exchange of protons behind the hydrodynamic slip plane for the larger sodium ions. This could, in principle, displace $CO_2$ and increase the local dielectric constant. Third, following [24] one may speculate about the adsorption of $Cl^-$ onto the $CO_2$ layer, forming some kind of outer Helmholtz plane. This should be addressed with experiments employing different co-ions. Again, simulations may shed some light on the underlying mechanisms of re-charging. More work is needed to explore this issue.

Finally, whatever the mechanism underlying the observed significant effects of dissolved $CO_2$, our findings underline the crucial importance of carefully monitoring and controlling the $CO_2$ content. Only in strictly $CO_2$ free samples, we find the theoretically expected salt concentration dependent decrease in electrophoretic mobility and $\zeta$-potential magnitudes. In the presence of $CO_2$, we observe the opposite. It thus appears that $CO_2$ issues may at least partially have caused the previously reported qualitative conflict between different studies.[19–23]

## CONCLUSION

We reported precision electrokinetic measurements of the charge state of flat and spherical dielectric surfaces in contact with aqueous electrolyte. We compared samples under strictly $CO_2$ free conditions to samples equilibrated at ambient air. The selective removal of $CO_2$ by ion exchange allows to attribute the observed effects to $CO_2$ and exclude other neutral gases as origin. We find that the presence of $CO_2$ reduces the surface charge. It also alters its salt concentration dependence of the surface charge. We note that the strong de-charging occurs for $CO_2$ concentrations as low as $6.3\,\mu mol\,L^{-1}$. It only became observable through the use of advanced conditioning techniques. In previous studies, much larger salt concentrations were typically employed, which masked this effect. Supported by a minimal model and by molecular simulations employing established force fields, we argue for the existence of a diffuse, dynamic layer of $CO_2$ at the dielectric surfaces. Further we suggest dielectric charge regulation to underlie the observed $CO_2$ dependent charge reduction. Our observations may shed light on previous conflicting reports on the salt-concentration-dependent electrokinetic behavior of dielectric surfaces. We anticipate that our study bears important consequences on future measurements in particular at low salt conditions as typical for some important applications. It further poses an interesting challenge to theory and simulation, to work out the mechanism(s) in detail. In any case, our study emphasizes the importance of a careful control of the experimental boundary conditions in experiments on surfaces in contact with water.

## ACKNOWLEDGMENT

Financial support of the DFG (Grant Nos. PA-459–15.2 and -19.1.) is gratefully acknowledged. This project has received funding from the European Research Council (ERC) under the European Union's Horizon 2020 research and innovation programme (grant agreement No 883631). N.M. acknowledges financial support by the Max Planck Graduate Center at the JGU, Mainz.

# Supporting Information

# Charging of dielectric surfaces in contact with aqueous electrolyte – the influence of $CO_2$

Peter Vogel[1], Nadir Möller[1], Muhammad Nawaz Qaisrani[1], Pravash Bista[2], Stefan A. L. Weber[2], Hans-Jürgen Butt[2], Benno Liebchen[3], Marialore Sulpizi[4] and Thomas Palberg[1]

[1] Institute of Physics, Johannes Gutenberg University, 55128 Mainz, Germany
[2] Max Planck Institute for Polymer Research, 55128 Mainz, Germany
[3] Institute for Condensed Matter Physics, Technische Universität Darmstadt, 64289 Darmstadt, Germany
[4] Department of Physics, Ruhr Universität Bochum, 44780 Bochum, Germany

# Table of contents

# 1 Experimental section

**Materials**

PnBAPS tracer particles are a 40:60 W/W copolymer of poly-n-butyl acrylamide (PnBA) and polystyrene (PS), kindly provided by BASF, Ludwigshafen (Lab code PnBAPS359, manufacturer Batch No. 2168/7390). Their diameter of $2a$ = 359 nm was determined by the manufacturer utilizing analytical ultracentrifugation.

Standard microscopy glass slides (75 × 25 × 1 mm, soda lime glass of hydrolytic class 3 by VWR International, Germany) served as charged wall specimen. They were sonicated prior to use for 30 min in 2% alkaline detergent water solution (Hellmanex III, Hellma Analytics), rinsed with double distilled water several times and dried with pressurized air. Low charge specimens were prepared by silanisation with perfluorooctadecyltrichlorosilane (PFOTS) in a chemical vapor deposition process. After oxygen plasma cleaning at 300 W for 10 min (Femto low-pressure plasma system, Diener electronic), the slides were placed in a vacuum desiccator containing a vial filled with 0.5 mL of 1H,1H,2H,2H-perfluorooctadecyltrichlorosilane (97%, Sigma Aldrich). The desiccator was evacuated to less than 100 mbar and the reaction was allowed to proceed for 30 min.[1]

**Sample conditioning**

**Tracer pre-conditioning**

By dilution with doubly distilled water, we prepared stock suspensions of approximately $n$ = 1×10$^{18}$m$^{-3}$, added mixed-bed ion exchange resin (IEX) (Amberjet, Carl Roth GmbH + Co. KG, Karlsruhe, Germany) and left it to stand under occasional stirring for some weeks. Then, the suspension was coarsely filtered using Sartorius 5 µm filters to remove dust, ion-exchange debris and coagulate regularly occurring upon first contact of suspension with IEX. All further conditioning was performed by circuit conditioning.

**Circuit-conditioning**

Pre-conditioned stock suspensions of tracers are diluted with double distilled water to the desired particle concentration. They are loaded into a peristaltically driven conditioning circuit under filtering with Sartorius 1.2 and 0.8 µm filters. All further sample preparation is performed in a closed system including the measuring cells and the preparation units.[2,3] Circuit preparation provides well-homogenized samples and allows precise adjustment of experimental boundary conditions. A schematic drawing is given in Figure S1a.

The suspension is pumped by a peristaltic pump (P) through a closed and gas tight Teflon® tubing system (grey arrows). Care is taken to assure a circuit free of $CO_2$-leaks. In particular, all components are equipped with gas tight tube fittings (Bohländer, Germany). The tubing connects i) the ion exchange chamber (IEX) filled with mixed bed ion exchange resin (Amberjet, Carl Roth GmbH + Co. KG, Karlsruhe, Germany). It can be bypassed, using the three-way valves (V) (Bohländer, Germany); ii) a reservoir (R) under inert gas atmosphere (Ar or $N_2$) to add suspension, water, or salt solutions; iii) a cell for conductivity measurements (C) and iv) several cells for the optical experiments (OC$_i$). One of these cells typically is a rectangular quartz-cell (5mm × 10mm) for turbidity measurements[4], to determine the actual particle number density from the transmitted intensity. Another is a rectangular cell (5mm × 10mm) for photometric pH measurements.[5] Further, we have the actual electrokinetic flow through cell with the flat specimen under study mounted as side walls.

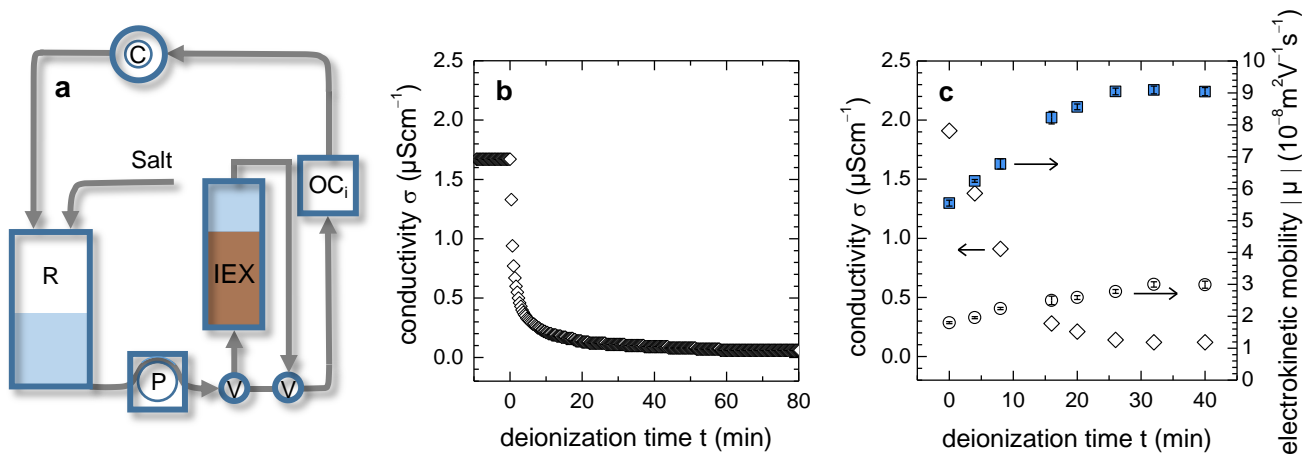

**Figure S1.** a) Schematic of the peristaltically driven conditioning circuit: R: reservoir, P: peristaltic pump, IEX: ion exchange column, C: conductometric cell, OC$_i$: optical (flow through) cell(s). All components are connected by gas tight tubing (grey lines). Respective flow directions are indicated as arrows. After deionization, the IEX can be bypassed using the valves (V). Salt solution can be added to the reservoir and its concentration monitored *via* conductivity. b) Evolution of the conductivity as a function of time under circuit conditioning for water. The water is continuously cycled. At $t$ = 0s it is set flowing through the IEX column. A low conductivity of about 0.2 µScm$^{-1}$ is already reached after 10 min of pumping. Complete removal of $CO_2$ takes about an hour. c) Development of conductivity (open diamonds) and electrophoretic mobility of a glass substrate (blue squares) and of PnBAPS359 with increasing conditioning time.

**Characterization of experimental boundary conditions**

pH is measured photometrically by adding small amounts of a standard pH indicator fluid (pH 3.0–10.0 pH indicator solution, Sigma-Aldrich).[5] From measured light intensity transmitted through either the sample or a water filled reference cell we calculated corresponding wavelength dependent absorbance values using Beer-Lambert law. Subsequently, absorbance ratios, $A_{\lambda 1}/A_{\lambda 2}$, were calculated from both absorbance maxima at $\lambda_1 = 430$ nm and $\lambda_2 = 615$ nm, respectively. For calibration of absorbance ratios, we used the indicator solution in buffer solutions adjusted to pH values between 3.0 and 8.0 in 0.5 pH unit steps. Absorbance spectra were recorded on a spectrophotometer (AvaSpec, Avantis) using low indicator dye concentrations by tenfold dilution of provided dye solutions from the manufacturer.

Conductivity is measured at a frequency of $\omega = 400$ Hz (electrodes LTA01 and LR325/01 with bridge LF538 or electrode LR325/001 with bridge LF340, WTW, Germany). Typically, the sample is cycled until a stable minimum conductivity is reached, defining the thoroughly deionized state. Low values are reached quickly. A stable minimum conductivity is obtained after about an hour. The progressing deionization for water is shown in Figure S1b. For water, we regularly obtain a minimum conductivity of 55–60 nScm$^{-1}$. This corresponds to the ion product of water: $c_{H+} \, c_{OH-} = 10^{-14}$ mol$^2$L$^{-2}$. The progressing deionization of a tracer suspension and the simultaneous increase in substrate electrophoretic mobility is shown in Figure S1c. In thoroughly deionized suspension, the final conductivity readings are enlarged by the particles and the amount of counter-ions released by the particles. The extra proton concentration, $c_{H+} = nZ_\sigma/(1000 N_A)$ amounts to a few µmol/L. Here, $n$ is the particle number density and $N_A$ is Avogadro's number. CO$_2$-free conditions are stable for a few hours, before small gas leaks lead to a noticeable rise in conductivity, respectively ion concentration.

This deionization process is fully reversible by adding desired amounts of salt solution and subsequently homogenize the tracer suspension by cycling under by-pass of the IEX column. To fully reverse de-carbonization, the deionized and de-carbonized suspension is brought into contact with ambient air (opening the reservoir lid) and cycled under bypassing the IEX column until the CO$_2$ dissolution and dissociation reactions have equilibrated. In a salt free suspension, the conductivity reads: $\sigma = n \, e \, Z_\sigma (\mu_{ep} + \mu_{H+}) + \sigma_B$.[6] Here, $e$ is the elementary charge; $\mu_{ep}$ and $\mu_{H+}$ are the mobilities of the tracers (independently measured) and of the protonic counter-ions, respectively; $n$ is the particle number density, which for calibration is determined from an independent experiment (turbidity). The background conductivity from water and – if applicable – equilibrated carbonic acid is denoted by $\sigma_B$. The obtained CO$_2$-concentrations are cross checked by pH measurements. Both agree well within experimental error and with the values obtained from calculations based on the free-ware program aqion.[7]

Experiments at elevated electrolyte concentration start from either the deionized or the CO$_2$-equilbrated state. Again, the IEX column is by-passed. Then, salt solution (Titrisol 0.1 molL$^{-1}$ NaCl or HCl, Merck, Germany) is added with a syringe through the septum of the reservoir in small quantities. The conductivity quickly reaches a constant larger value.

To vary the CO$_2$-concentration in a controlled way, we start from a thoroughly deionized and de-carbonized state. We then let the system get into contact with ambient air by opening a reservoir valve for a short time. Subsequently, the system is cycled again until a stable conductivity reading is reached. Now, however, the system is cycled under bypassing the IEX column. Subsequently the system is de-ionized again. To obtain a larger CO$_2$ concentration a longer contact time is allowed for. If the reservoir valve is left open under continued cycling, the system becomes fully equilibrated with ambient CO$_2$ within a few minutes. The obtained state of re-carbonization is monitored by parallel conductivity and pH measurements. From these, the actual CO$_2$ concentrations can be inferred, either from the conductivity readings using Hessinger's model[6] or from the pH readings using the free-ware program aqion.[7] CO$_2$-concentrations obtained from both approaches typically agree well within experimental error.

**Total micro-ion concentration**

The two result plots Figure 2 and Figure 3 show data in dependence on the total concentration of micro-ions, to allow comparison of the three different electrolyte types in dependence on a single control-parameter. This quantity was calculated respecting the contributions of all micro-ionic species in solution, i.e., the tracer particle counter-ions, the added electrolyte ions, and the OH$^-$ ions stemming from water hydrolysis. The contribution of wall-counter-ions is not resolvable in bulk conductivity measurements and was, therefore, neglected.

Upon the addition of NaCl to CO$_2$-free systems of known tracer density (Figure 2a and c and Figure 3), $n$, the measured pH stays at its value close to neutrality, as set by the small amount of tracer counter-ions. The concentration of OH$^-$ is calculated via the ion product of water. The amount of added anions and cations is calculated point by point from the added volumina of salt solution. The result is controlled *via* conductivity readings evaluated using Hessinger's model of independent ion migration.[6] For their evaluation, we obtained the number of freely mobile tracer counter-ions, $Z_\sigma$, from the tracer density dependent conductivity measurements. We explicitly note that upon exchanging non-mobile H$^+$ tracer counter-ions for Na$^+$ counter-ions the electrolyte content does not change, it is nevertheless clearly visible in the conductivity measurements. In fact, upon adding NaCl, the conductivity initially increases like expected for the addition of HCl, and only later increases like expected for NaCl, after the exchange is completed. This is well described by our conductivity model. For the micro-ion limiting conductivities, the known literature values were used. The measured small change in tracer mobility, $\mu_{ep}$ could not be resolved within the experimental uncertainty of the conductivity experiments, and, in view of the several times larger proton mobility, was therefore neglected. We obtained very good consistency of the total micro-ion content as calculated by summing up the known inputs and out conductivity evaluations.

For the addition of NaCl to fully CO$_2$-equilibrated systems (Figure 2b and d), we assumed that without observed pH changes, also the composition of CO$_2$ related reaction products of Equation (1) does not change. We therefore simply added the amount of 6.3µmol$^{-1}$ as a background of ions stemming from CO$_2$-reactions. Interestingly, after CO$_2$-equilibration, the conductivity increases strictly linearly with the amount of added salt. Moreover, we find the increase to correspond to that expected for NaCl. This indicates, that in the presence of $CO_2$, there is no counter-ion exchange. The number of mobile counter-ions, i.e., the effective conductivity charge had decreased significantly (see also below, 1.3.1). Respecting these $CO_2$ specific changes, we again found good consistency between total electrolyte concentrations based on the added amounts of salt solution and the corresponding conductivity readings.

For the addition of HCl to $CO_2$-free systems (Figure 3), the measured pH decreases with increasing HCl concentration. In comparing measured values to theoretical expectations for the added quantities, we respected the small amount of $H^+$ already present in the form of tracer counter-ions. With this, we found good agreement between measured and calculated pH values. In this case, the conductivity increased linearly as a function of added HCl and with the expected slope. In the absence of counter-ion exchange, the system behaved simply additive. Again, good consistency for the total micro-ion concentration was observed from the different approaches.

Finally, starting from a $CO_2$- and salt-free system, we increased the concentration of dissolved $CO_2$ in steps, as described above. In this case, the respective amounts of $CO_2$ added are not known *a priori*. In the absence of micro-ionic species other than those shown in Equation (1), we could therefore resort to the pH-readings to calculate the ionic concentrations using the aqion software.[7] In each case, the amount of formed carbonate ($CO_3^{2-}$) was extremely small as compared to the amounts of $H^+$ and $CO_3H^-$. To these values, we added the tracer counter-ions. Since no data for the effective tracer charge were available (e.g., from NaCl concentration dependent conductivity experiments at known intermediate $CO_2$ concentrations) we approximated the number of tracer counter-ions by the electrokinetic tracer charge number. The latter was derived from the measured mobilities using the aqion results for $H^+$ as input to calculate the Debye screening parameter. Thus, for the addition of $CO_2$ we summed over the concentrations of $H^+$, $OH^-$, and $CO_3H^-$ in calculating the total micro-ionic concentration.

The small amounts of divalent anions are not discernible in the conductivity measurements, despite their larger valence. Nor was there a significant deviation due to the lowered tracer mobility. However, the $CO_2$-induced de-charging was clearly seen *via* the decreasing number of tracer $H^+$ counter-ions with their high mobility. The latter effect was therefore accounted for in the point-by-point evaluation of the corresponding conductivity readings. Again, an excellent agreement was observed between pH-based calculations and conductivity-based control.

It thus was possible to present the data of Figure 2 and 3 consistently in dependence on a reliably determined, single control parameter. A nice side effect was, that measurements at the same total micro-ionic concentration performed for HCl and $CO_2$ also were done at the same pH. In the literature, many data on pH-dependent $\zeta$-potentials can be found. To allow a comparison for the interested readers, we display the pH-dependent magnitudes of $\zeta$-potentials for $SiO_2$ and PnBAPS in Figure S2a and b. Data are redrawn from Figure 3a and b of the main text, where they were initially displayed in dependence on total micro-ion concentration.

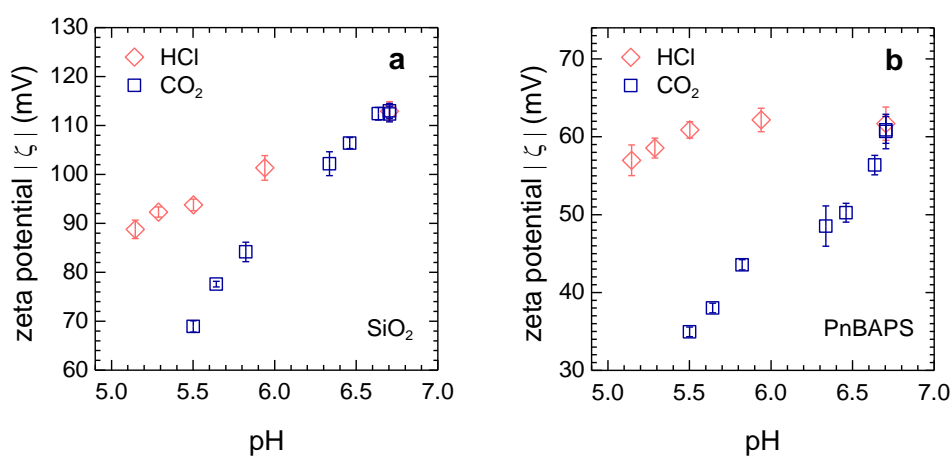

**Figure S2**: Magnitudes of $\zeta$-potentials in dependence on photometrically obtained pH values. Data are for HCl (upper curve, red diamonds) and $CO_2$ (lower curve, blue squares). a) Data for $SiO_2$ (redrawn from Fig. 3a), and b) Data for PnBAPS (redrawn from Fig. 3b). Note that these data were taken starting from the $CO_2$ - and salt-free state. Then small amounts of either HCl solution or gaseous $CO_2$ were added, and the systems equilibrated by cycling under bypass of the IEX. The initial pH (right side) is slightly below 7, due to the presence of particle counter-ions ($H^+$). The final pH of 5.5 is reached after extended contact with ambient air.

**Tracer and wall characterization**

**Tracer concentration**

The volume fraction, $\Phi$, of the stock suspension of PnBAPS359 was obtained from drying and weighing experiments using the known density of PnBAPS of 1.05 gcm$^{-3}$. The number density was estimated using $n = 3\Phi/(4\pi a^3)$ with $a$ denoting the known particle radius. For turbidity measurements on further diluted samples we followed Botin.[4,8] Samples were equilibrated with ambient $CO_2$ and observed in the rectangular flow-through quartz cuvettes of $2d$ = 5mm optical path length (Hellma, Germany). The transmittance, $T = I(n)/I_0$, was recorded at $\lambda$ = 532.5 nm as function of $n$. Here, $I_0$ denotes the transmitted intensity for cuvettes filled with doubly distilled water. In Figure S3a, data are plotted in terms of attenuation, $A$, per millimetre with $A = 10 \times \log(I_0/I)$ (dB). At low concentrations, the suspensions are disordered. Their static structure factor S(q) = 1. Then, the transmitted intensity obeys the Beer-Lambert law, $2dn\sigma_{532.5} = -\ln(I/I_0)$.[9] From fits to the data, we obtained an attenuation cross section of $\sigma_{532.5} = (3.2 \pm 0.1) \times 10^{-2}$ µm$^2$. This allows an online determination of the number density in our electrokinetic experiments. Typically, $n$ = 1.5 × 10$^{16}$ m$^{-3}$, which is well within the range of linear attenuation.

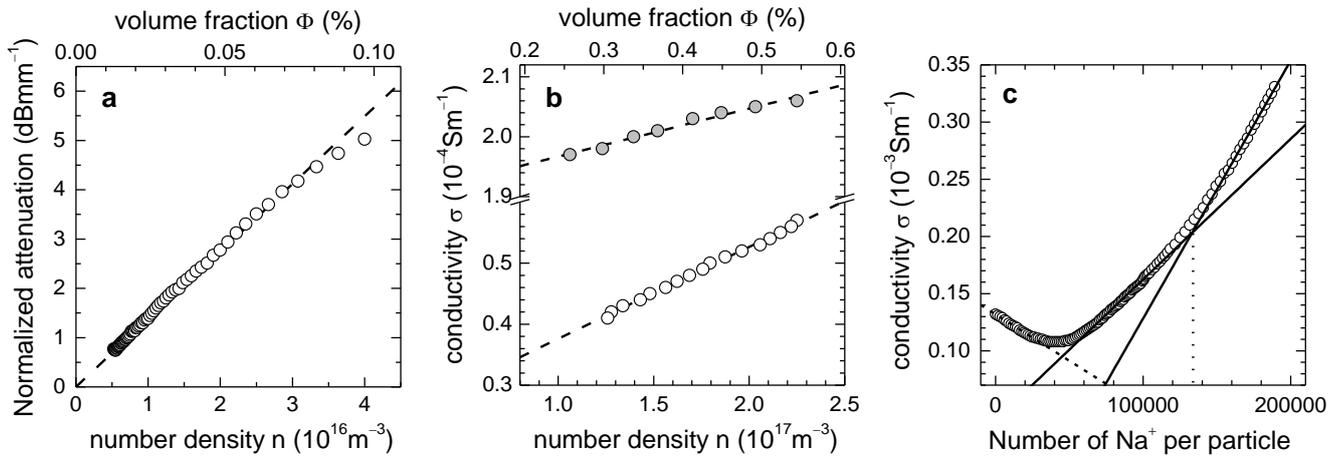

**Figure S3.** Optical and charge characterization of PnBAPS359 tracer particles. a) Attenuation of transmitted intensity at $\lambda$ = 532.5nm normalized to the cell thickness, $2d$ = 5mm, as a function of number density $n$ (lower scales) and volume fraction $\Phi$ (upper scales), respectively. The dashed line is a fit of the Beer-Lambert attenuation law returning an attenuation cross section of $\sigma_{532.5} = (3.2 \pm 0.1) \times 10^{-2}$ µm$^2$. b) Concentration dependence of the low frequency conductivity of salt-free PnBAPS359 with (filled circles) and without $CO_2$ (open circles). Dashed lines are fits of the independent ion migration model[6] returning $Z_\sigma = -(2350 \pm 35)$ for the $CO_2$-free case and $Z_\sigma = -(1300 \pm 65)$ after $CO_2$-eqilibration. c) Titration curve for the addition of NaOH starting from thoroughly deionized PnBAPS359. Conductivity data are shown as function of the number of Na$^+$ added per particle. The solid lines are linear fits to the data returning the number of ionogenic surface groups, $N = (13 \pm 1) \times 10^4$, from the second equivalence point at elevated amount of added Na$^+$.

**Tracer charge**

Conductivity experiments on salt-free suspensions of PnBAPS359 in dependence on $n$ yield the effective number of freely mobile counter-ions. Results are shown in Figure S3b for two cases, with and without $CO_2$. The dashed lines are linear least squares fits to the data using $\sigma = n\,e\,Z_\sigma(\mu_{ep} + \mu_{H+}) + \sigma_B$.[6] For thoroughly deionized $CO_2$-free systems, we obtain the effective conductivity charge as $Z_\sigma = -(2350 \pm 35)$. For realistic salt free systems, we find $Z_\sigma = -(1300 \pm 65)$. These numbers have to be compared to the number of dissociable groups. Figure S3c shows the results of a conductivity titration starting from deionized/decarbonated PnBAPS359. The titration curve is typical for the simultaneous presence of strong and weak acid groups.[6,10,11] From the crossing points of linear fits before and past the second equivalence point, we have a titrated group number $N = (13 \pm 1) \times 10^4$. From the initial slope we extract a charge ratio of $Z_\sigma/Z \approx 0.12$, and from that a bare charge $Z$ of $Z \approx 1.9 \times 10^4$. As expected and in agreement with previous work,[12,13] the conductivity effective charge is much smaller than the bare charge, which in turn is much smaller than the group number $Z_\sigma \ll Z \ll N$. Moreover, the effective charge is considerable smaller than the bare charge, due to charge renormalization.

Dividing the particle surface area by the group number, $N$, yields a mean physical parking area per surface group of $A_0 \approx 3$ nm$^2$. The average spacing between ionogenic groups is approximately 1.7 nm $\approx 2.5\,\lambda_B$. The parking area for dissociated groups under $CO_2$ free conditions is 0.5 nm$^2$. Thus, mutual electrostatic interactions are not expected to interfere with dissociation. However, even without $CO_2$-induced effects, there is only weak degree of dissociation related to the group p$K$ and the low pH within the innermost part of the double layer.

**Flat substrate contact angles**

For contact angle measurements advancing and receding contact angles were determined with sessile water drops (Contact angle measurement system, OCA35, Dataphysics). An 8 µL water drop was deposited on the surface. Then 16 µL of deionized water was pumped into the drop and subsequently sucked out at a rate of 0.5 µL/s by a Hamilton syringe connected to a hydrophobic needle. The process was repeated three times without interruption. During inflation and deflation, drops were imaged in side-view. Contact angles were calculated by fitting an ellipse model to the contour images.[1,14] The advancing and receding contact angles of PFOTS coated slides, measured with a sessile drop, were (120 ± 2)° and (87 ± 2)°, respectively. The contact angles of bare glass slides treated according to the cleaning procedure described in section 1.1 were (22 ± 2)°.

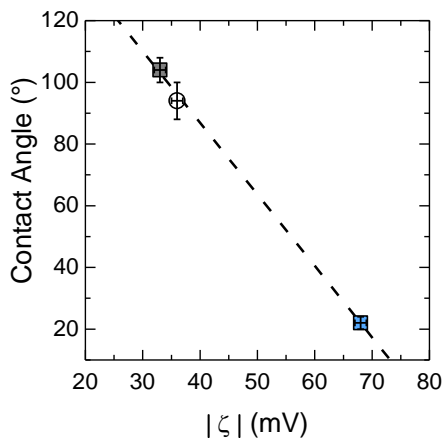

**Figure S4.** Correlation of contact angles and magnitudes of the $\zeta$-potentials for PFOTS (grey squares), PS (open circles) and glass (blue squares) surfaces, both measured under ambient conditions.

**Custom-made electrokinetic cell**

Electrokinetic mobilities were determined for both tracers and substrates in the same electrokinetic cell. With the cell connected to the conditioning circuit, this ensures identical experimental boundary conditions. The flow-through cell was custom made of Poly-Methyl-Methacrylate (PMMA) (Figure S5a).[15] It features a rectangular cross-section with large aspect of $K = h/d = 20$, i.e. cell length and cell height, $h$, are much larger than cell depth, $d$. The conditioned standard microscopy glass slides (75 × 25 × 1 mm) serving as charged wall specimen are mounted as windows.

Platinized platinum electrodes with 14/20 ground glass joints (Rank Bros. Bottisham, UK) are inserted into the electrode chamber of the cell. The effective electrode distance $L = 12$ cm was calibrated from conductivity measurements using potassium chloride electrolyte solutions. Data in electrokinetic measurements are recorded with an electric field applied. Figure S5a shows the field direction with the anode to the right and the cathode to the left. It also shows the coordinate system used in interpretation.

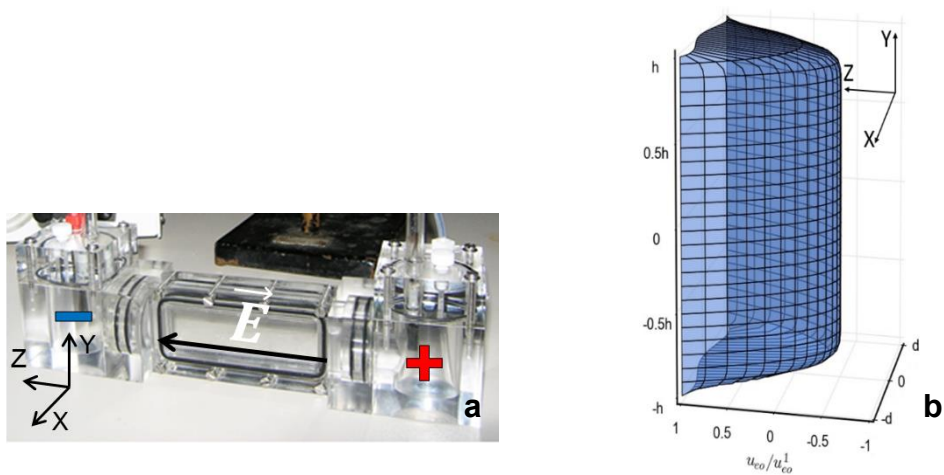

**Figure S5.** a) Custom-made electrokinetic flow-through cell with exchangeable side walls. Positions of electrodes inserted into the electrode chambers and applied electric field **E** pointing in $z$-direction are indicated. The sample conditioning circuit (Figure S1a) is connected at each electrode chamber. b) Calculated parabolic flow profile in the electrokinetic cell shown in a). The $x$, $y$, $z$-axis represents the normalized cell depth $d$, normalized cell height $h$, and the solvent velocity normalized to the electro-osmotic velocity at the wall, respectively.

Presence of charged walls leads to pronounced electro-osmotic flow towards the cathode. In closed cells this induces a solvent backflow in the cell centre. At each location, the measured tracer particle velocity, $\mathbf{v}_p = \mathbf{v}_{ep} + \mathbf{v}_s(\mathbf{v}_{eo}, x, y)$, is the unique sum of its constant electrophoretic velocity, $\mathbf{v}_{ep} = \mu_{ep}\mathbf{E}$, and the position dependent solvent velocity, $\mathbf{v}_s(x,y)$, depending on the electroosmotic velocity, $\mathbf{v}_{eo} = \mu_{eo}\mathbf{E}$, at the cell wall. The flow profile in Figure S5b was calculated according to White[16] using the boundary condition of no net solvent flow and $K = 20$. In our experiments we measure the tracer velocity distribution at mid-plane cell height ($y = 0$). Under this conditions the parabola-like particle flow profile at mid cell height ($y = 0$) can be analytically approximated as:[17,18]

$$\mathbf{v}(x, y=0) = \mu_{ep}\mathbf{E} + \mu_{eo}\mathbf{E}\left[1 - 3\left(\frac{1 - \frac{x^2}{d^2}}{2 - \frac{384}{\pi^5 K}}\right)\right] \quad \text{(S1)}$$

The cell is integrated into the conditioning circuit during conditioning. It can be sealed from it by electromagnetic valves during measurements. An alternating square-wave field of strength up to $E_{MAX} = U/L = 15$ V cm$^{-1}$ is applied. To avoid accumulation of tracers at the electrodes and to further ensure fully developed stationary flows[19], field switching frequencies $f_{AC} = (0.02 - 0.1)$ Hz were used. Measurement intervals were restricted to one field direction. Each was starting after the full development of electro-osmotic flow profile and ending shortly before the field reversal. For each field strength, we recorded and averaged $250 - 400$ full time-frame power spectra.

**Laser Doppler Velocimetry**

We recently introduced a Super-Heterodyne (SH) variant of standard heterodyne laser Doppler velocimetry (LDV). A scheme of Super-Heterodyne Laser Doppler Velocimetry (SH-LDV) is shown in Figure S6.

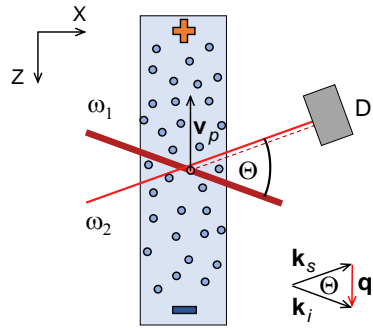

**Figure S6.** Scattering scheme with the optical paths of illuminating ($\omega_1$) and reference ($\omega_2$) beams and definition of scattering vector $\mathbf{q}$ (mid plane cut, $y = 0$). The direction of particle velocity $\mathbf{v}_p$ is oppositely to the applied electric field direction as indicated by the arrow.

The instrument used here features a small scattering angle and an integral cross sample detection. It was described earlier in much detail together with the corresponding scattering theory.[4,15] Two laser beams cross in the sample cell containing colloidal tracers moving at a velocity $\mathbf{v}_p$ in an electric field, $\mathbf{E}$, applied in z-direction. The scattering vector $\mathbf{q} = \mathbf{k}_i - \mathbf{k}_s$ (where $\mathbf{k}_i$ and $\mathbf{k}_s$ are the wave vectors of the illuminating and scattered beams) is collinear to the applied electric field direction. Its modulus is given by $q = (4\pi\nu/\lambda_0) \sin(\Theta/2)$ with suspension refractive index $\nu$, laser wavelength $\lambda_0 = 632.8$nm and $\Theta = 8°$ the beam crossing angle in the sample cell. At each particle, light of the illumination beam of frequency $\omega_1$ is scattered towards the detector (D) under an angle $\Theta$. It is Doppler shifted by $\Delta\omega = \mathbf{q}\cdot\mathbf{v}_p$. Here, $\mathbf{q}$ is the scattering vector of magnitude: $q = (4\pi\nu/\lambda_0) \sin(\Theta/2)$ with suspension refractive index $\nu$ and laser wavelength $\lambda_0$. Light scattered in Figure S6 by PnBAPS tracers moving upwards in negative z-direction is positively Doppler shifted. Hence, a positive Doppler shift corresponds to negatively charged particles moving with a negative velocity with respect to the field direction. At the detector, scattered light of frequency $\omega = \omega_1 + \Delta\omega$ mixes with the reference beam of frequency $\omega_2 = \omega_1 + \omega_{SH}$. The resulting beat signal is analyzed by a frequency analyzer (OnoSokki DS3000, Compumess, Germany) to obtain the power or Doppler spectrum $C_{SH}(\mathbf{q},\omega)$, being the time Fourier-transformation of the mixed-field intensity autocorrelation function, $C_{SH}(\mathbf{q},\tau)$:[20]

$$C_{SH}(\mathbf{q},\omega) = \frac{1}{\pi}\int_0^{+\infty} d\tau \cos(\omega\tau) C_{SH}(\mathbf{q},\tau) \quad \text{(S2)}$$

where $\omega$ denotes the circular frequency and $\tau$ the correlation time.

We first consider the case of strictly local detection. We assume the particles to undergo Brownian motion with an effective, electric field dependent diffusion coefficient, $D_{eff}$, and directed motion with a single, constant drift velocity indicated by the index 0. We exploit the statistical independence of diffusion and drift velocity. The corresponding power spectrum reads:

$$C_{SH}^0(\mathbf{q},\omega) = \left(I_r + \langle I_s(\mathbf{q})\rangle\right)^2 \delta(\omega)$$
$$+ \frac{\langle I_s(\mathbf{q})\rangle^2}{\pi} \frac{2\mathbf{q}^2 D_{eff}}{\omega^2 + (2\mathbf{q}^2 D_{eff})^2}$$
$$+ \frac{I_r \langle I_s(\mathbf{q})\rangle}{\pi}\left[\frac{\mathbf{q}^2 D_{eff}}{(\omega + [\omega_{SH} + \Delta\omega])^2 + (\mathbf{q}^2 D_{eff})^2} + \frac{\mathbf{q}^2 D_{eff}}{(\omega - [\omega_{SH} + \Delta\omega])^2 + (\mathbf{q}^2 D_{eff})^2}\right] \quad (S3)$$

where $I_r$ is the reference beam intensity and $\langle I_s(\mathbf{q})\rangle$ the time-averaged scattering intensity in dependence on $\mathbf{q}$.

The spectrum contains three contributions. First, a static background term $\delta(\omega)$ centred at zero frequency. Second, a homodyne Lorentzian of width $\Gamma = 2\mathbf{q}^2 D_{eff}$, centred at the origin and independent of drift velocity. This term is well known from standard dynamic light scattering. Third, two super-heterodyne Lorentzians of spectral width $\Gamma = \mathbf{q}^2 D_{eff}$ shifted symmetrically away from the origin by $\omega_{A,B} = \pm(\omega_{SH} + \mathbf{q}\cdot\mathbf{v}_p)$, i.e. both by $\omega_{SH}$ and the Doppler shift frequency $\Delta\omega = \mathbf{q}\cdot\mathbf{v}_p$. Note that it is this additional super-heterodyning frequency shift of the reference beam by $\omega_{SH}$, which allows shifting the SH-term away from any low-frequency noise and separate it from homodyne scattering of the second term. The desired information regarding electrophoretic and diffusive particle motion is fully contained via $\Delta\omega$ in each of the two super-heterodyne Lorentzians. It is thus sufficient to analyse measured data only for positive frequencies.[4]

In the present set-up, scattered light is collected from an extended detection volume comprising the full $x,z$-cross section at mid cell height ($y = 0$). The observable particle velocity, $\mathbf{v}_p$, depends on position (c.f. Equation S1), leading to a distribution of particle velocities and Doppler frequencies, respectively. The spectral shape of the selectively detected SH-part then results from a convolution of the distribution $p(\Delta\omega) = p(\mathbf{q}\cdot\mathbf{v}_p)$ with a Lorentzian of width $\Gamma = \mathbf{q}^2 D_{eff}$.[15,20]

$$C_{SH}(\mathbf{q},\omega) = \int d(\mathbf{q}\cdot\mathbf{v}_p)\, p(\mathbf{q}\cdot\mathbf{v}_p)\, C_{SH}^0(\mathbf{q},\omega) \quad (S4)$$

The super-heterodyne signal thus averages over all particle velocities and Doppler frequencies, respectively, present in the scattering volume. We note, that for a given field strength, the shape of $C_{SH}(\mathbf{q},\omega)$ is solely determined by the solvent flow profile whereas its centre of mass is determined by the particle electrophoretic velocity and its broadening by the effective diffusion coefficient. Thus, we have three statistically independent variables for the fitting of the spectra. For fitting and evaluation we follow the procedures outlined in much detail in previous work.[4,15,20,21]

Figure S7a shows a set of spectra recorded at different field strengths in cells with bare glass side walls at salt- and $CO_2$-free conditions. The SH-frequency of $f_{SH} = \omega_{SH}/2\pi = 3$kHz corresponds to zero particle velocity. The characteristic spectral shape corresponds to a parabola like velocity profile.[17] With increasing field strength, the spectra stretch and their centre of mass shifts towards larger Doppler frequencies. The solid lines are least square fits of our model (Equation S1 used in Eqns. S2–S4) returning three independent parameters: $\mathbf{v}_{ep}$ from the spectral center of mass, $\mathbf{v}_{eo}$ from the field dependent width, and the tracer diffusivity, $D_{eff}$, from the homogeneous broadening. Our particles and walls are negatively charged, resulting in negative velocities and mobilities. Figure S7b shows the moduli of extracted velocities to increase linearly with field strength. Least squares fits of a linear function (dashed lines) return mobilities of $\mu_{ep} = \mu_{PnBAPS} = -(3.1\pm0.2)\times10^{-8}$ m$^2$V$^{-1}$s$^{-1}$ and $\mu_{eo} = \mu_{SiO2} = -(9.0\pm0.1)\times10^{-8}$ m$^2$V$^{-1}$s$^{-1}$, respectively. The quoted uncertainties correspond to the standard errors of the linear fits at a confidence level of 0.95. The effective tracer diffusivity is extracted from the spectral width $\Gamma$ of the super-heterodyne Lorentzian and increases linearly with increasing field strength (see Figure S7c). As discussed in Ref.[20], this can be related to Taylor dispersion in the parabolic solvent flow.

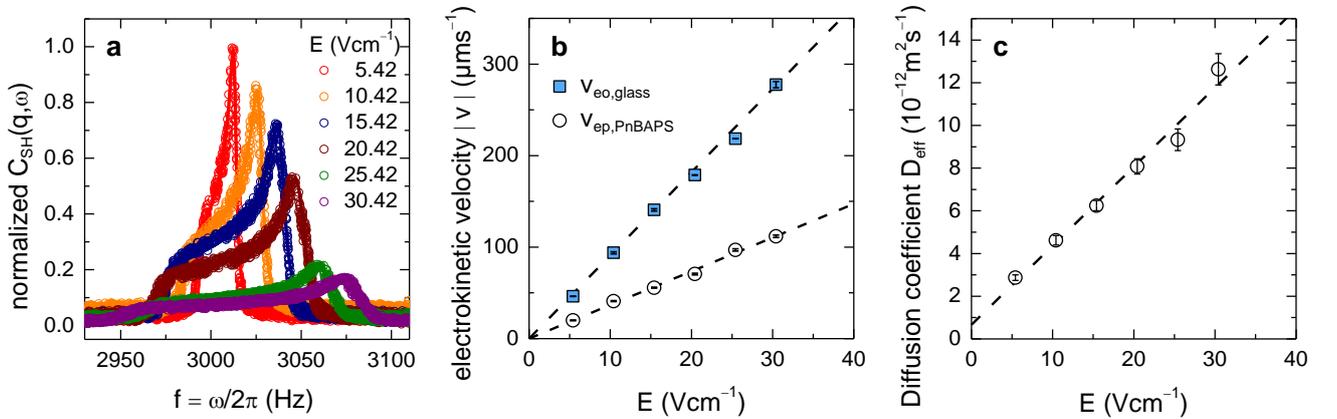

**Figure S7.** a) Doppler Spectra (symbols) recorded on PnBAPS tracer particles in a cell with bare glass walls at salt- and $CO_2$-free conditions for different field strengths as indicated. Solid lines are least square fits to the data based on Eqn. S1 – S4. b) The field dependence of both electro-kinetic velocities shows linear behaviour. A least square fit (dashed line) returns electro-kinetic mobilities of $\mu_{ep,PnBAPS} = -(3.1\pm0.2)\times10^{-8}$m$^2$V$^{-1}$s$^{-1}$ and $\mu_{eo,SiO2} = -(9.0\pm0.1)\times10^{-8}$m$^2$V$^{-1}$s$^{-1}$. c) Effective tracer diffusivity plotted versus $E$ corresponding to field strength dependent spectral broadening.

**Mobility evaluation**

We base our evaluation for $\zeta$-potentials on the recommendations of the International Union of Pure and Applied Chemistry (IUPAC), which provide suitable solutions of the electrokinetic equations and standard protocols for different experimental boundary conditions.[22] In its latest extensions, the underlying mean field description includes effects of retardation, double layer polarization, surface conductance, ion-adsorption, electrical double layer overlap and related charge regulation.[23,24]

To be specific, we use the protocol of O'Brien and White[25] but consider realistic salt free conditions[21,23,26], i.e., we rely on the general electrokinetic model for concentrated suspensions in aqueous electrolyte solutions, which predicts electrophoretic mobility and electrical conductivity in static electric fields under realistic conditions.[23] Most specifically, we explicitly account for *all* ionic species, $i$, present in solution. We consider released counterions (H$^+$) by dissociation of particle surface groups, H$^+$ and OH$^-$ from water hydrolysis, H$^+$ and bicarbonate anions, HCO$_3^-$, from dissolved ambient CO$_2$ and ions from salination (NaCl or HCl). The Debye-Hückel screening parameter reads: $\kappa = \sqrt{e^2 / \varepsilon_0 \varepsilon_r k_B T \sum_i n_i z_i}$, where $n_i$ and $z_i$ are the micro ionic number densities and valences, respectively. The total electrolyte concentration is $c_s = \sum_i c_i = \sum_i n_i /(1000 N_A)$, respectively $\sum_i c_i = n_{ci}/(1000 N_A) + 2c_{H_2O} + 2c_{NaCl} + 2\sqrt{K_a c_{CO_2}}$, where $N_A$ is Avogadro's number, $n_{ci} = Z_\sigma n$ the counter ion number density, $c_{H_2O}$ the concentration of ions from water autoprotolysis, $c_{NaCl}$ the ion concentration from added monovalent salt and $K_a$ the effective equilibrium dissociation constant of carbonic acid ($pK_a = 6.3$ [7,27]). In the simplest case of an infinitely extended flat, charged surface, we have thin double layers ($\kappa a \gg 1$) and $\mu = \varepsilon_0 \varepsilon_r \zeta / \eta_s$ where $\eta_s$ is the solvent viscosity.[28] For spherical surfaces, retardation and double layer relaxation introduce a dependence on $\kappa a$.[25,29] Here, we use the radius determined from ultracentrifugation.

Observed mobilities are converted to reduced mobilities $\mu_{red} = 3\mu\eta_s e/2\varepsilon_0\varepsilon_r k_B T$, to eliminate the influence of solvent viscosity, dielectric permittivity, and temperature.[30] We here used $\varepsilon_{water} = \varepsilon_0\varepsilon_r = 7.17\times10^{-10}$ AsV$^{-1}$m$^{-1}$ and $\eta_{water}$ (T = 25°C) = 0.891×10$^{-3}$ Pas. In theoretical work, predictions of mobility magnitudes are often plotted for different constant $\kappa a$ values versus reduced $\zeta$-potential magnitudes, $\zeta_{red} = \zeta e/(k_B T)$, e.g., in Ref.[25]. This is adequate for spheres as long as no change of $\kappa$ occurs. Alternatively, both observed mobility magnitudes and mobility magnitudes predicted for constant reduced $\zeta$-potential magnitudes can be plotted versus $\kappa a$ (cf. Figure S8).[26,29,31] In Figure S8, we show this for PnBAPS359.

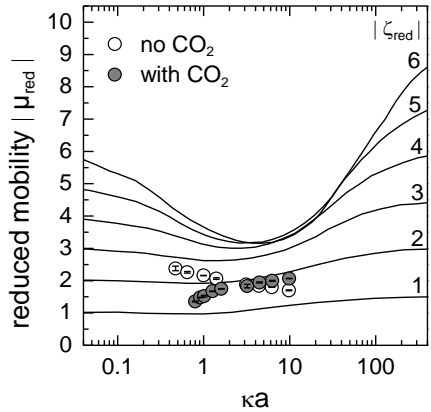

**Figure S8.** Magnitudes of reduced tracer mobilities $\mu_{red}$ plotted against reduced screening parameter $\kappa a$ obtained for the salt and CO$_2$-free reference state (open circles) and upon CO$_2$ equilibration in contact with ambient air (grey filled circles). Solid lines represent theoretical predictions according to standard electrokinetic theory for realistic conditions at respective constant reduced $\zeta$-potential magnitudes.

## Theoretical section

### Analytic Theory

Here we develop a minimal model to show that a $CO_2$ enrichment layer is energetically feasible at all, i.e., that a conceivable effective wall-$CO_2$ interaction strength (with a maximum strength of $k_BT$ or even less) leads to a $CO_2$ enrichment layer with a thickness which is comparable to the size of the $CO_2$ molecules. For the present system of interacting neutral $CO_2$ molecules at an attractive dielectric wall in aqueous environment, we consider the $N$-particle Smoluchowski-equation for the joint probability density $P(\mathbf{r}_1,\mathbf{r}_2,...,\mathbf{r}_N)$ for particles with an isotropic pair interaction potential $V = (1/2)\sum_{i,j=1}^{N} U(r_{ij})$, where $r_{ij} = |\mathbf{r}_i - \mathbf{r}_j|$, in an external wall potential $V_{ext} = \sum_{i=1}^{N} U_{ext}(z_i)$:[32]

$$\dot{P} = \sum_{i,j=1}^{N} \nabla_i D_{ij}\left[\beta\left[\nabla_j(V+V_{ext})\right]P + \nabla_j P\right] \tag{S5}$$

where $D_{ij}$ are the elements of the microscopic diffusion matrix and $1/\beta = k_BT$ is the thermal energy. To model the aggregation of interacting particles at an attractive wall, we are interested in the density distribution under steady state conditions. Thus, we set $\dot{P} = 0$ such that the square bracket on the right-hand side has to varnish. We then reduce equation (S5) to an equation for the one-particle probability density given by $P(\mathbf{r}) = \int\int...\int d\mathbf{r}_2\, d\mathbf{r}_3...d\mathbf{r}_N\, P(\mathbf{r},\mathbf{r}_2,...,\mathbf{r}_N)$.

Applying the mean field approximation, $P(\mathbf{r},\mathbf{r}') \approx P(\mathbf{r})P(\mathbf{r}')$, after performing the truncated Taylor expansion $P(\mathbf{r}') \approx P(\mathbf{r}) - [\nabla P(\mathbf{r})] \cdot (\mathbf{r}-\mathbf{r}')$ and using $V_{ext}(\mathbf{r}) = V_{ext}(z)$, we obtain the following equation for the (macroscopic) density $\rho(\mathbf{r}) = N P(\mathbf{r}) \approx (N-1) P(\mathbf{r})$, where for symmetry reasons $\rho(\mathbf{r}) = \rho(z)$:

$$-\frac{1}{\beta}\frac{d}{dz}\ln\rho + \frac{4\pi}{3}\left(\frac{d}{dz}\rho\right)\int_{r_{min}}^{\infty} dr\, r^3 \frac{d}{dr}U(r) - \frac{d}{dz}U_{ext} = 0 \tag{S6}$$

Here we have replaced the lower boundary of the integration range by a minimal cutoff distance $r_{min}$ accounting for the nearest possible distance between two particles.

Now defining $I = \frac{4\pi}{3}\int_{r_{min}}^{\infty} dr\, r^3 \frac{d}{dr}U(r)$, this equation reduces to

$$-\frac{1}{\beta}\frac{d}{dz}\rho + I\rho\frac{d}{dz}\rho - \rho\frac{d}{dz}U_{ext} = 0 \tag{S7}$$

Specifying the external wall potential to $U_{ext}(z) = -(f_0/3)/z^3$ and choosing the boundary condition as $\rho(z\to\infty) = \rho_\infty$, the solution of equation (S7) reads

$$\rho^*(z^*) = \frac{1}{\alpha^*}W\left(\alpha^* \exp\left(\alpha^* + \frac{1}{z^{*3}}\right)\right) \tag{S8}$$

where $W$ is the product logarithm, and where we have defined the reduced density as $\rho^* = \rho/\rho_\infty$ and the reduced length as $z^* = z/\xi$. $\xi$ is the characteristic length scale given by

$$\xi = \left(\frac{f_0}{3k_BT}\right)^{1/3} \tag{S9}$$

We note that the solution (S8) depends on one dimensionless parameter only, which measures the relevance of the interactions

$$\alpha^* = -\frac{I\rho_\infty}{k_BT}. \tag{S10}$$

In the absence of pair-interactions, i.e., $\alpha^* = 0$, the solution (S8) reduces to

$$\rho^*(z^*) = \exp\left(\frac{1}{z^{*3}}\right) \tag{S11}$$

which translates into $\rho(z) = \rho_\infty \exp(-\beta U_{ext}(z))$ when reinstating the dimensional units.

**Parameters**

For numerical calculations, the saturation concentration of ambient $CO_2$ dissolved in water was assumed to be $\rho_\infty = 1.18 \times 10^{-5}$ mol/L $\approx 7.1 \times 10^{21}$ m$^{-3}$ and the thermal energy was estimated to be $k_B T \approx 3.6 \times 10^{-21}$ J. Interpreting the particle pair interaction potential as Lennard-Jones potential, $U(r) = U_{LJ}(r) = 4\varepsilon\left[(\sigma/r)^{12} - (\sigma/r)^6\right]$, the distance at which $U_{LJ}$ vanishes was approximated to be $\sigma_{CO_2} \approx 0.37$ nm. If the attractive part of the interactions among $CO_2$-molecules is due to Van der Waals interactions we have $U_{VdW}(r) = -C/r^6 = -4\varepsilon\sigma^6/r^6$ yielding $\varepsilon = C/(4\sigma^6)$, where $C$ is the coefficient of the Van der Waals interactions of two $CO_2$ molecules in water. We assume $\varepsilon \approx k_B T/2$ for the effective LJ-interaction coefficient of the $CO_2 - CO_2$ interaction in water.[33] This corresponds to a Van der Waals interaction coefficient $C = 4\sigma^6\varepsilon \approx 1.8 \times 10^{-77}$ Jm$^6$ which would be a typical particle interaction strength in vacuum. Let's assume that we have the same $\varepsilon$ for the interaction between wall molecules and $CO_2$ molecules. Now using $U_{ext}(z) = -\pi n_{glass} C/(6z^3)$ we have $f_0/3 = (\pi/6) n_{glass} C \approx 2.5 \times 10^{-49}$ Jm$^3$ with $n_{glass} \approx 2.65 \times 10^{28}$ m$^{-3}$.

Next, we estimate values of $I$. We obtain for the minimal interparticle distance $r_{min} = \delta\sigma$:

$$I = \frac{4\pi}{3}\int_{\delta\sigma}^{\infty} dr\, r^3 \frac{d}{dr} U_{LJ}(r) = \frac{4\pi}{3}\left[\frac{8(3\delta^6 - 2)}{3\delta^9}\right]\sigma^3\varepsilon \tag{S12}$$

That is, the value of $I$ is highly sensitive on the lower cutoff of the integral (minimal possible distance). Choosing e.g., $\delta = 1/2$ we obtain $I = -(1.12 \times 10^4)\sigma^3\varepsilon$ and for $\delta = 3/4$ we obtain $I \approx -218\sigma^3\varepsilon$ (for $\delta = 0.9$ we have $I \approx -11.7\sigma^3\varepsilon$). Thus, for $\varepsilon = k_B T/2$ and $r_{min} = \delta\sigma = 3\sigma/4$ we have $\alpha^* \approx 3.9 \times 10^{-5}$ and for $r_{min} = \sigma/2$ we have $\alpha^* \approx 2.0 \times 10^{-3}$ showing a strong increase in the importance of interparticle interactions when reducing the minimal distance between particles about a factor of 1.5.

For estimating the value of the characteristic length scale $\xi$, from the above, we obtain $f_0/(3k_B T) \approx 6.9 \times 10^{-29}$ m$^3$ which yields according to equation (S9)

$$\xi \approx 0.41 \text{ nm} \tag{S13}$$

which approximately equals $\sigma_{CO_2}$. That is, a particle – wall distance of $z = \sigma_{CO_2}$ corresponds to $z^* = z/\xi \approx 1$ and $\rho \approx 2\rho_\infty$. We note that at a wall distance of $z = \xi/2$ we have $\rho(z^*=1/2) \approx 2 \times 10^3 \rho_\infty$. Thus, decreasing the wall distance about factor 2 increases the $CO_2$ density about factor $10^3$.

**Asymptotics**

Near the wall, i.e., for small, reduced length, $z^* \ll 1$, we obtain for the reduced density:

$$\rho^*(z^*) = \frac{1}{\alpha^* z^{*3}} + h.O. \tag{S14}$$

This power law is due to the $CO_2$ cross-interactions. Without interactions no such power law occurs. Far away from the wall, i.e., for large, reduced length, $z^* \gg 1$ (where the $CO_2$ density is low), the ideal-gas result ($\alpha^* = 0$) provides an excellent fit:

$$\rho^*(z^*) = \exp\left(\frac{1}{z^{*3}}\right) \tag{S15}$$

**Conclusions and limitations of the minimal model**

The functional form of the $CO_2$ density profile, $\rho^*$, depends (in mean-field) only on one effective (dimensionless) parameter, $\alpha^*$, which does not involve the $CO_2$-wall-interactions. However, the strength of the latter interaction controls the length scale, $\xi$, over which the $CO_2$ density is significantly enhanced near the wall. Notably, we find that this length scale increases with the $CO_2$ – wall interaction coefficient to the power of 1/3, i.e., it is not very sensitive on the precise value of the coefficient. Based on our results we can conclude that as long as the effective LJ-interaction coefficient, $\varepsilon$, between wall-molecules and gas-molecules is on the order of $\approx 1 k_B T$ the present minimal theory supports the idea that one layer of significantly enhanced $CO_2$ density is present at the wall (but probably not more than one layer). It cannot, however, make any assertions about the effects of hydrophobicity, charging and competition with water structure formation close to the interface.

**Molecular Dynamics Simulations**

**Computational Details**

To investigate the $CO_2$ adsorption at the silica/water interface, we simulated different models of silica surfaces with different degree of hydrophilicity/hydrophobicity selected from Heinz's minerals models database.[34] The silica slabs with variable surface chemistry and an adjustable degree of ionization were created as supercells from the unit cell of the $\alpha$-quartz (Q2 environment) and $\alpha$-cristobalite (Q4, Q3 environments) crystal structures. The Q4 surface is uncharged and silanol-free and mimics more a hydrophobic interface compared to the regular Q3 surface which presents a total silanol group density up to 4.7 $nm^{-2}$. For more details on Heinz's silica models and their surface chemistry, we will refer the reader to the reference [34]. In this preliminary work, we explore the structural behaviour of $CO_2$ at Q4, and dissociated Q3 surfaces as two model cases of hydrophobic and hydrophilic interfacial systems without and with charge. A more complete study exploring the $CO_2$ adsorption in a larger variety of chemical environments will be the subject to upcoming work. This will also include implementation of charge regulation by Coulombic silanol-silanol interactions.

All the initial slab structures were placed in the centre of the simulation box with a single face exposed to the solvent. The box dimensions for the silica Q4 system were 3.168nm, 3.308nm, and 16.368nm while for Q3 the cell geometries are 3.340nm, 3.487nm, and 13.371nm, respectively. In all cases, 50 molecules of $CO_2$ were randomly added to the simulation box and finally, the box was filled with the water molecules. The interactions between the silica atoms are described by the force field from the work of Emami et. al.[34] Ions interactions were taken into account by the standard Charmm27 force field, while $CO_2$ parameters were taken from reference [35]. Water was described by TIP3P model.[36]

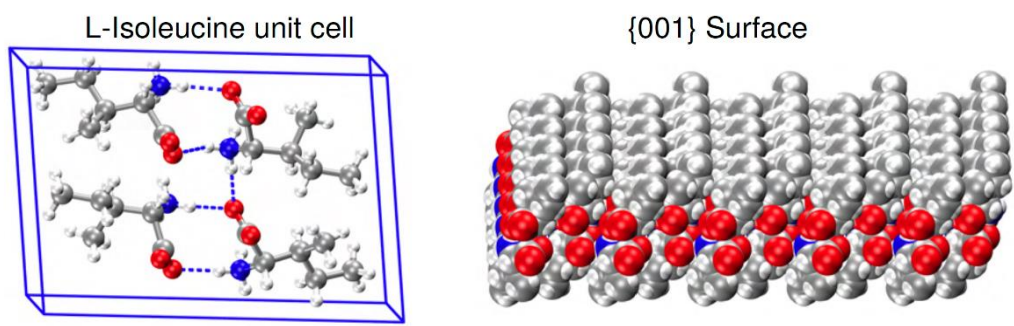

**Figure S9.** Snapshots of the unit cell of L-isoleucine crystal and its 001 plane.

The hydrophobic organic surface was created as a supercell from the experimentally resolved L-isoleucine crystal.[37] Isoleucine residues contain a hydrophobic side chain consisting of $CH_2$ and $CH_3$ groups. The {001} plane of the unit cell consists of all the $CH_3$ groups exposed to the solvent. The crystal structure of L-isoleucine is held together by the strong hydrogen bonds formed between the charged terminal groups, as well as by the van der Waals forces acting between the hydrophobic sidechains. Figure S9 shows the unit cell of the L-isoleucine crystal and its {001} plane. For this system, we used the standard Charmm27 force field [38,39] together with TIP3P water model.

All simulations were carried out in the canonical ensemble, where the velocity-rescaling thermostat[40] with the relaxation time of 0.1 ps, was used to control the temperature. A time step of 1 fs was set to update the atom's positions and velocities. In all simulations, three-dimensional periodic boundary conditions were applied. For non-bonded interactions, a pair list was created with a cut-off radius of 1.4 nm and updated after every 10 fs. The cut-off for the shifted Lennard-Jones potential was set at 1.2 nm. For the long-range electrostatic interactions, the Particle Mesh Ewald-Switch[41] method was used with a Coulomb switching cut-off 1.2 nm. To truncate the Van-der-Waals interactions, a dispersion correction was applied. A short 1 ns semi-isotropic NPT simulation was performed to equilibrate the solvent density. The pressure control was carried out by using Parrinello-Rahman barostat.[42] After the equilibration, all the production simulations were extended up to 50 ns.

## Results

At the hydrophobic isoleucine/water interface strong $CO_2$ adsorption is observed. In the $CO_2$ density profile (Figure S10, right panel) a peak can be observed directly located at the solid surface, which excludes water from the first adsorbed layer. In the case of the silica Q4 surfaces, we also observed strong $CO_2$ adsorption as shown in the right panel of Figure S11. In this case, the $CO_2$ density peak position coincides with the water density first position indicating a competition between $CO_2$ and water in the first adsorbed layer. This can be seen in the left panel of Figure S11.

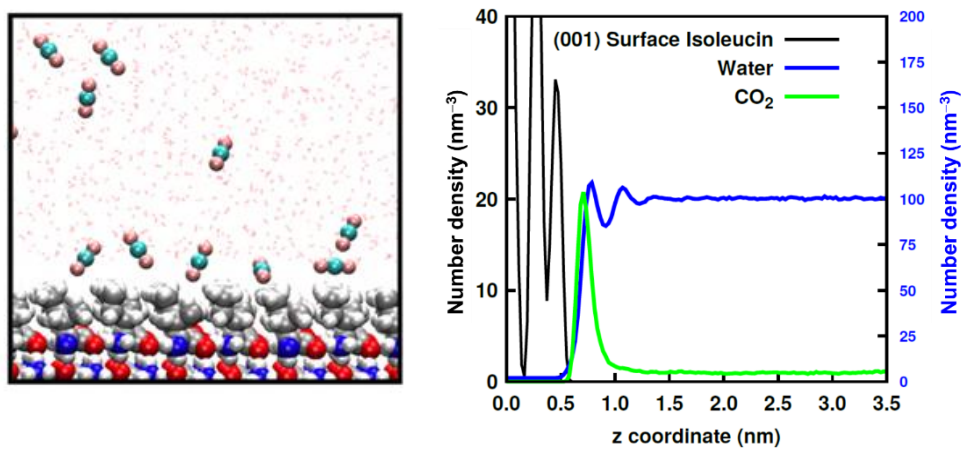

**Figure S10:** Left panel: A random snapshot is taken from the MD simulation of the Isoleucine surface in contact with water and $CO_2$. The right panel is showing the density profile of water (blue) and $CO_2$ (green). The scale for the $CO_2$ (green) density is reported on the left y-axis, while the scale for the water density (blue) is reported on the right y-axis.

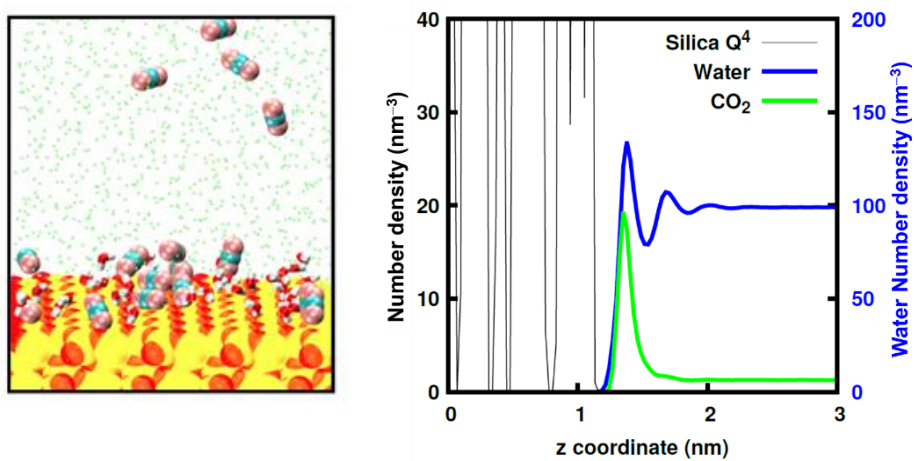

**Figure S11**: Left panel: A random snapshot from molecular dynamics simulations showing competing water and $CO_2$ layer at a neutral silica Q4 surface (without silanol groups). The right panel illustrates the density profile of water (blue, right vertical axis), $CO_2$ (green, left vertical axis) and silica Q4 surface (black, right vertical axis).

Increasing the silanol density on the surface, from 0 to 2.4 nm$^{-2}$ and 4.7 nm$^{-2}$ (but still staying charge neutral) reduces the intensity of the $CO_2$ peak at the silica/water interface. Note that the experimentally determined density of ionisable groups was 0.33 nm$^{-2}$ from titration. It is interesting to notice that even in the case of the largest silanol density the $CO_2$ still shows a significant preference for the interface with respect to the bulk. This is shown in Figure S12.

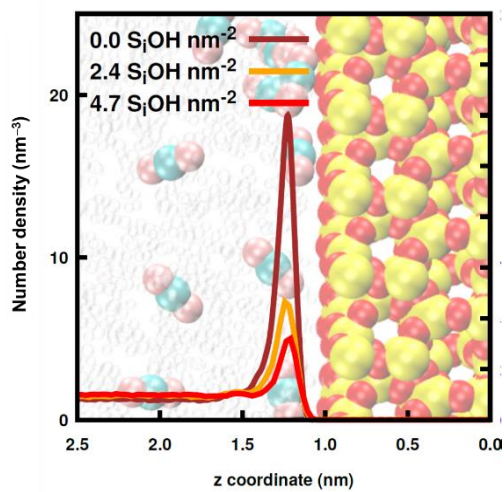

**Figure S12**: $CO_2$ number density at the silica Q4 surface (brown line) and at the silica Q3 surface with silanol group density 2.4 nm$^{-2}$ (orange line) and 4.7 nm$^{-2}$ (red line) respectively.

Finally, moderate preference for the interface persists even at a mild degree of surface deprotonation, e.g., up to 9% for a silica Q3 surface carrying a silanol density of 4.7 nm$^{-2}$ (See Figure S13 and Figure 4b of the main text). This is a parking area of 2.3 nm$^2$ per charge or an average distance 1.53 nm between charges. At these small distances we would already anticipate a significant amount of charge regulation under conditions of effectively reduced relative dielectric constants. A larger degree of protonation starting from 18% and above would instead lead to $CO_2$ depletion at the interface. The left panel shows a snapshot from a simulation run at 5% dissociation detailing the rich interfacial structure. This snapshot is also employed in the TOC.

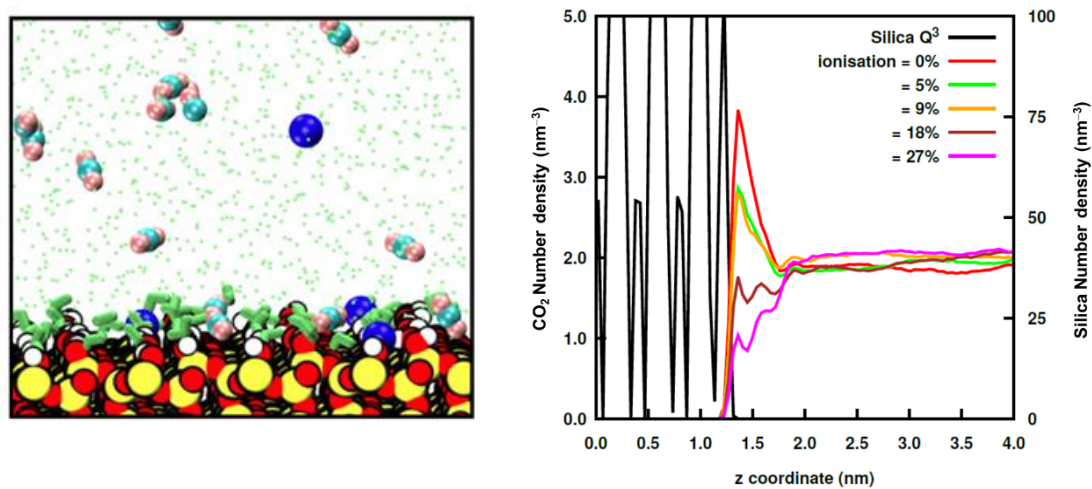

**Figure S13**: Left panel: A random snapshot taken from the MD simulations of the 5% dissociated Q3 surface. Blue spheres are the counter sodium ions, dots in lime colour represent water for clarity while at the surface the first hydration layer is shown in lime sticks representation. $CO_2$ molecules are shown in a combination of cyan(carbon) and pink(oxygen) colour. Right panel: Density profile of $CO_2$ at Q3 surface with a silanol density 4.7 nm$^{-2}$ with different degrees of ionisation.